\documentclass[article]{aastex61}
\usepackage{hyperref}
\usepackage{ textcomp }
\usepackage[]{natbib}

\newcommand{\etal}{{et.~al}}
\newcommand{\kms}{{km~s$^{-1}$}}
\newcommand{\sst}{{\it Spitzer Space Telescope}}

\begin{document}

\title{Infrared Photometric Properties of 709 Candidate Stellar Bowshock Nebulae}

\author{Henry A. Kobulnicky}
\author{Henry A. Kobulnicky}
\author{Danielle P. Schurhammer}
\author{Daniel J. Baldwin }
\author{William T. Chick}
\affiliation{Department of Physics \& Astronomy, University of Wyoming, Dept 3905, Laramie, WY 82070-1000, USA}
\author{Don M. Dixon}
\author{Daniel Lee}
\author{Matthew S. Povich}
\affiliation{Department of Physics \& Astronomy, California State Polytechnic University, 3801 West Temple Avenue,  Pomona, CA 91768, USA}

\begin{abstract}

Arcuate infrared nebulae are ubiquitous throughout the
Galactic Plane and are candidates for partial shells,
bubbles, or bowshocks produced by massive runaway stars.  We
tabulate infrared photometry for 709 such  objects using
images from the {\it Spitzer Space Telescope (SST)}, {\it 
Wide-Field Infrared Explorer (WISE)}, and {\it Herschel
Space Observatory (HSO)}.  Of the 709 objects identified at 
24 or 22 $\mu$m, 422 are detected at the $HSO$ 70 $\mu$m 
bandpass. Of these, only 39 are detected at $HSO$ 160 
$\mu$m.   The 70 $\mu$m peak surface brightnesses are
0.5---2.5 Jy  $arcmin^{-2}$. Color temperatures calculated
from the 24  $\mu$m to 70  $\mu$m ratios range from 80~K to
400~K. Color temperatures from 70 $\mu$m to  160 $\mu$m
ratios are systematically lower, 40~K to 200~K.  Both of
these temperature are, on average, 75\% higher than the 
nominal temperatures derived by assuming that dust is in
steady-state radiative equilibrium.   This may be evidence
of stellar wind bowshocks sweeping up and heating ---
possibly fragmenting but not destroying --- interstellar
dust.   Infrared luminosity correlates with standoff
distance, $R_0$, as predicted by published hydrodynamical
models.  Infrared spectral energy distributions are
consistent with interstellar dust exposed to a either single
radiant energy density, $U=10^3$---$10^5$ (in more than half
of the objects) or a range of radiant energy densities
$U_{min}$=25 to  $U_{max}$=10$^3$---10$^5$ times the mean
interstellar value for the remainder.  Hence, the central OB
stars dominate the energetics, making these enticing
laboratories for testing dust models in constrained
radiation environments. SEDs are consistent with PAH
fractions $q_{PAH}\lesssim$1\% in most objects. 

\end{abstract}

\keywords{Catalogs ---
Stars: massive --- 
Interstellar medium (ISM), nebulae --- 
surveys --- 
(ISM:) HII regions ---
(Stars:) early-type 
}

\section{Introduction} \label{sec:intro}

Infrared nebulae exhibiting distinctive circular,
elliptical, or arc-like morphologies reveal a host of
astrophysical phenomena sculpting the Galactic interstellar
medium.   Massive stars or star clusters may blow wind- and
photon-driven  bubbles \citep{Castor1975, Weaver1977}, which
may grow to galactic scales if with sufficient energy
injection \citep{Heiles1979, MacLow1988}. Individual massive
stars or evolved stars ejecting  may carve smaller bubbles
on $<$few parsec scales. 
\citet{Churchwell2006,Churchwell2007} compiled a catalog of
nearly 600 interstellar bubbles discovered in mid-infrared 
3.6--8.0 $\mu$m images of the Galactic Plane obtained during
the \sst\ Galactic Legacy Infrared MidPlane Survey
Extraordinaire program
\citep[GLIMPSE;][]{Benjamin2003,Churchwell2009} conducted
with the  Infrared Array Camera \citep[IRAC;][]{Fazio2004}. 
Worldwide volunteers identified hundreds more  using these
images as part of the Zooniverse Milky Way Project
\citep{Simpson2012} citizen science project.  This wealth of
identifications has led to the discovery of new star
clusters, evolved stars, and better understanding of the
structure and evolution of the Milky Way. Bubbles may
develop an asymmetric, elliptical, or arcuate appearance in
the presence of a density gradient in  the ambient medium or
a relative motion between the energy source and ISM; in the
latter case a stellar bowshock nebula is  formed
\citep{Wilkin1996,Wilkin2000}.

Infrared nebulae with arcuate morphologies may reveal the
presence of a high-velocity massive star when the $>$1000
\kms\ stellar wind shocks the ambient ISM, generating a
distinctive bowshock feature.  \citet{Gull1979} studied the
bowshock (first seen in H$\alpha$) preceding the
high-velocity O9.2IV star $\zeta$ Oph.  
\citet{Brown2005b} discovered eight H$\alpha$ bowshocks
using modern H$\alpha$ sky surveys. However, H$\alpha$,
X-ray, and radio free-free emission are expected to be
several orders of magnitude fainter than the far-infrared 
dust emission, making the latter the most observable
bowshock signature \citep{Meyer2014,Meyer2016}.
\citet{vanBuren1988} and \citet{vanBuren1995} cataloged
dozens of far-infrared bright bowshocks associated with
known massive and high-proper-motion stars thought to be
``runaways'' \citep{Blaauw1961}.  Deliberate searches for
bowshock nebulae preceding high-velocity stars yielded a few
dozen more candidates \citep[e.g.,][]{Gvaramadze2008,
Kobulnicky2010,  Gvaramadze2013,Peri2012,Sexton2015}. 
Hypervelocity stars \citep{Hills1988,Kenyon2008}  populate
the extreme end of the runaway spectrum, but such objects
may not support bowshocks \citep{Brown2005,Meyer2014}.
Bowshocks have also been associated with high-mass X-ray
binaries \citep{Gvaramadze2011B}, pulsars \citep{Wang2013},
red \citep{Noriega1997, Gvaramadze2014A} and blue
\citep{Gvaramadze2014B} supergiants, and the A-type star
$\delta$ Vel \citep{Gaspar2008}.   Thus, arcuate nebulae
appear to be associated with different of types of
systems and different physical  phenomena. 
\citet{Peri2012} and \citet{Peri2015} concluded that only
10--15\% of  high-proper motion massive stars showed
evidence of bowshocks in {\it Wide Field Infrared Explorer}
\citep[$WISE$;][]{Wright2010} mid-infrared  images.  The
accumulating evidence suggests that early type stars and
high space velocities are neither necessary nor sufficient
conditions for the formation of prominent bowshocks at
infrared wavelengths.   \citet{Huthoff2002} compared the
velocities and ambient interstellar densities of OB stars
producing bowshocks  and those lacking bowshocks, concluding
that interstellar density  likely plays a larger role than
space velocity or stellar spectral type in creating  an
observable nebula. This conclusion is supported by
hydrodynamical simulations of \citet{Comeron1998} and
citet{Meyer2016} exploring how a range a range of stellar
and ISM properties combine to produce a variety of bowshock
morphologies or no visible nebula at all. Mildly supersonic
space velocities, strong stellar winds,  and high ambient 
interstellar densities $n>$0.1~cm$^{-3}$  result in the most
pronounced bowshock nebulae.  

\citet{Kobulnicky2016} conducted a extensive visual
examination covering hundreds of square degrees within the
infrared Galactic Plane surveys made by the $SST$ GLIMPSE 
and MIPSGAL \citep{Carey2008} programs, and other wide-area
$SST$ programs including  the $Spitzer$ Legacy Survey of the
Cygnus-X Complex \citep[][]{Beerer2010}, and the $Spitzer$
Mapping of the Outer Galaxy \citep[SMOG;][]{Carey2008}. 
Data from the  four-band all-sky $WISE$ survey within
several degrees of the Plane was also used to cover portions
not observed by $SST$.  They tabulated 709 arcuate nebulae
discovered either at the $SST$  24 $\mu$m or $WISE$ 22
$\mu$m bandpasses along with positional and photometric data
on the central prominent star located along the symmetry
axis of each object.   Figure 2 of
\citet{Kobulnicky2016}  plots the spatial distribution of
the objects along the Galactic Plane. In about 15--25\% of
the objects the nebulae was located within or pointed
toward  a nearby \ion{H}{2} region, leading the authors to
designate a class of ``in-situ'' bowshock candidates
following \citet{Povich2008} where a localized flow of
ambient of material impinges upon an otherwise  low-velocity
massive star.  \citet{Chick2017}  obtained optical spectra
for $\sim$60 of the central stars and found that  85\% were
O- or early B type stars, strengthening the evidence that
these nebulae are frequently bowshocks associated with hot
stars and likely constitute genuine bowshocks.

In this contribution we present aperture photometry  and
spectral energy distributions for the 709 bowshock
candidates cataloged by \citet{Kobulnicky2016}.  Their
infrared spectral energy distributions (SEDs) and derived
dust temperatures provide insight regarding the heating
mechanisms of the nebulae (i.e., shocks versus radiative
heating).  SEDs can also be used to infer the dust grain
size distribution and presence or absence of polycyclic
aromatic hydrocarbons (PAHS).     Section 2 describes he
photometry procedure employed to measure fluxes from the 3.6
$\mu$m $SST$ or 3.4 $\mu$m $WISE$  bandpasses through the 
{\it Herschel Space Observatory} Photoconductor Array Camera
and Spectrometer \citep[PACS;][]{Poglitsch2010}   70 and 160
$\mu$m bands.  This section includes two tables containing
the photometric measurements  available in machine-readable
format.  Section 3 presents an analysis of the photometric
properties, infrared colors, inferred dust temperatures. 
Section 4 presents an anlysis of spectral energy
distributions in comparison to  interstellar dust models. 
Section 5 compare the sizes and luminosities of the nebulae
to published hydronamical simulations of bowshock nebulae,
yielding some inferences concerning the densities of the
ambient interstellar medium.  

\section{Aperture Photometry}

Datasets for aperture photometry of 709 infrared nebulae
included the $SST$ 3.6, 4.5, 5.8, and 8.0  $\mu$m  (i.e., the IRAC
I1, I2, I3, I4 bands) mosaics and 24 $\mu$m  
Multiband Imaging Photometer for Spitzer
\citep[MIPS;][]{Rieke2004} produced by the GLIMPSE team
covering the GLIMPSE, MIPSGAL, SMOG, and Cyg-X survey
regions.  The telescope beam size at these bands is
1.66\arcsec\, 1.72\arcsec\, 1.88\arcsec\,  1.98\arcsec\ and
6\arcsec\  FWHM, 
respectively\footnote{http://irsa.ipac.caltech.edu/data/SPITZER/docs/irac/iracinstrumenthandbook,
\\ 
http://irsa.ipac.caltech.edu/data/SPITZER/docs/mips/mipsinstrumenthandbook}. 
For targets in the second and third Galactic quadrants not covered by $SST$
programs we use the all-sky $WISE$ atlas images at 3.4, 4.6,
12, and 22 $\mu$m (i.e., W1, W2, W3, W4 bands) which have
beamsizes of 6.1\arcsec, 6.4\arcsec, 6.5\arcsec, and
12.0\arcsec\ FWHM, respectively.  The Hi-Gal \citep[Herschel
Infrared Galactic Plane Survey;][]{Molinari2016}  covered
the entire Plane with the PACS instrument at 70 and 160
$\mu$m for which the beamsizes are approximately 6\arcsec\
and 12\arcsec, respectively.  Additional $HSO$ pointed
observations of individual objects outside the Plane (e.g.,
the prototypical bowshock runaway star $\zeta$ Oph
\citep{Gull1979}) were sometimes available.    We  assembled
the level 2.5 or higher data processed and mosaicked
($\sim$2$\times$2 degree UNIMAP images) from the  Infrared
Science Archive
(IRSA)\footnote{http://irsa.ipac.caltech.edu/frontpage/ }
for analysis. For the level of photometric precision
relevant to  this analysis, the different mapping algorithms
(e.g., UNIMAP verus MADmap) and data products available are
indistinguishable.   

\begin{figure}
\plotone{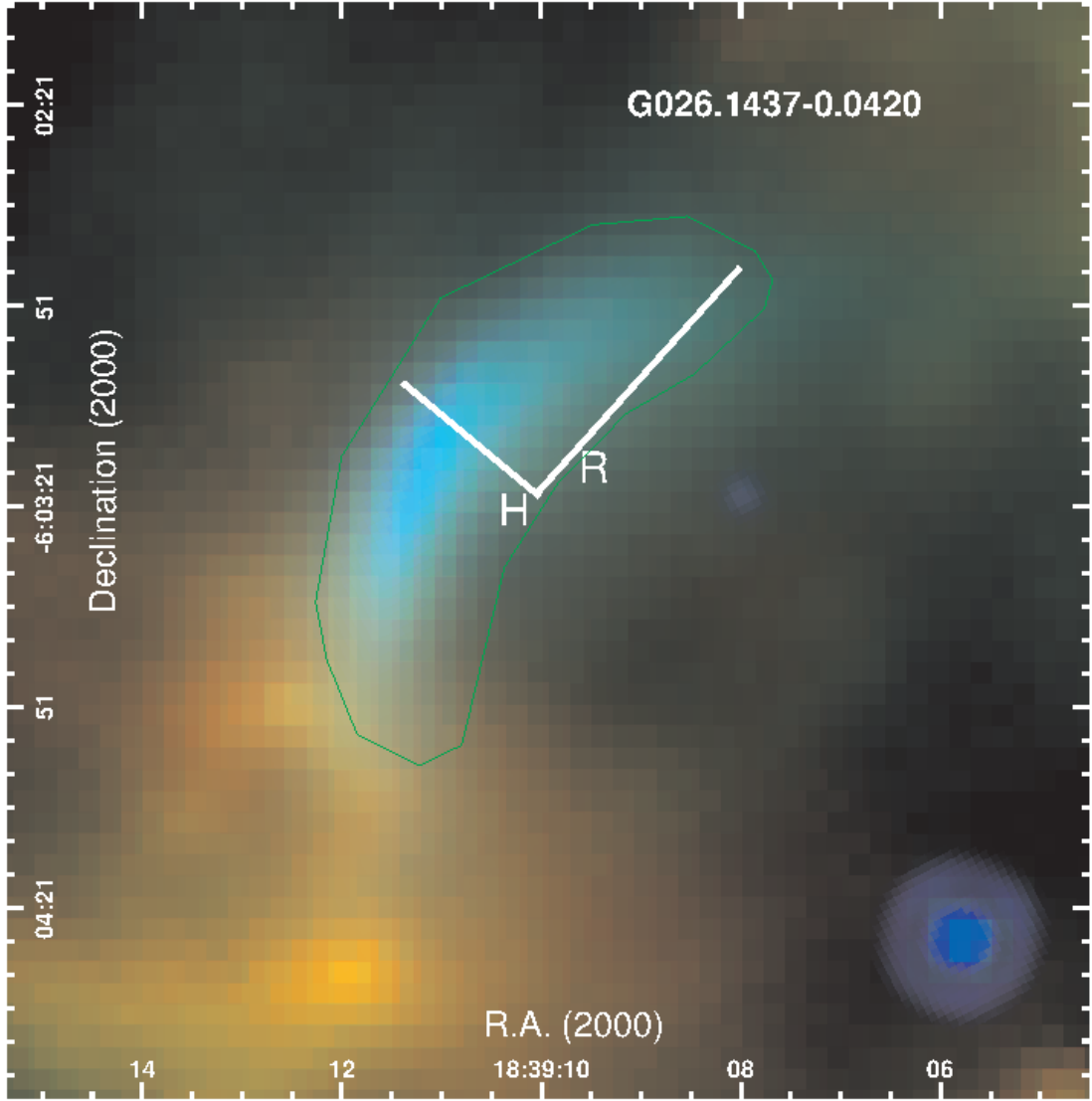}
\caption{Three-color representation of the G026.1473$-$0.0420 nebula (object \#123
from the catalog of \citet{Kobulnicky2016}) with blue/green/red representing 24/70/160 $\mu$m
from $SST$/$HSO$/$HSO$, respectively.   
The green polygon depicts the irregular aperture used to conduct aperture photometry.
White bars depicts a nominal angular Height and Radius assigned to this object.  
\label{fig:example1}}
\end{figure}

\begin{figure}
\plotone{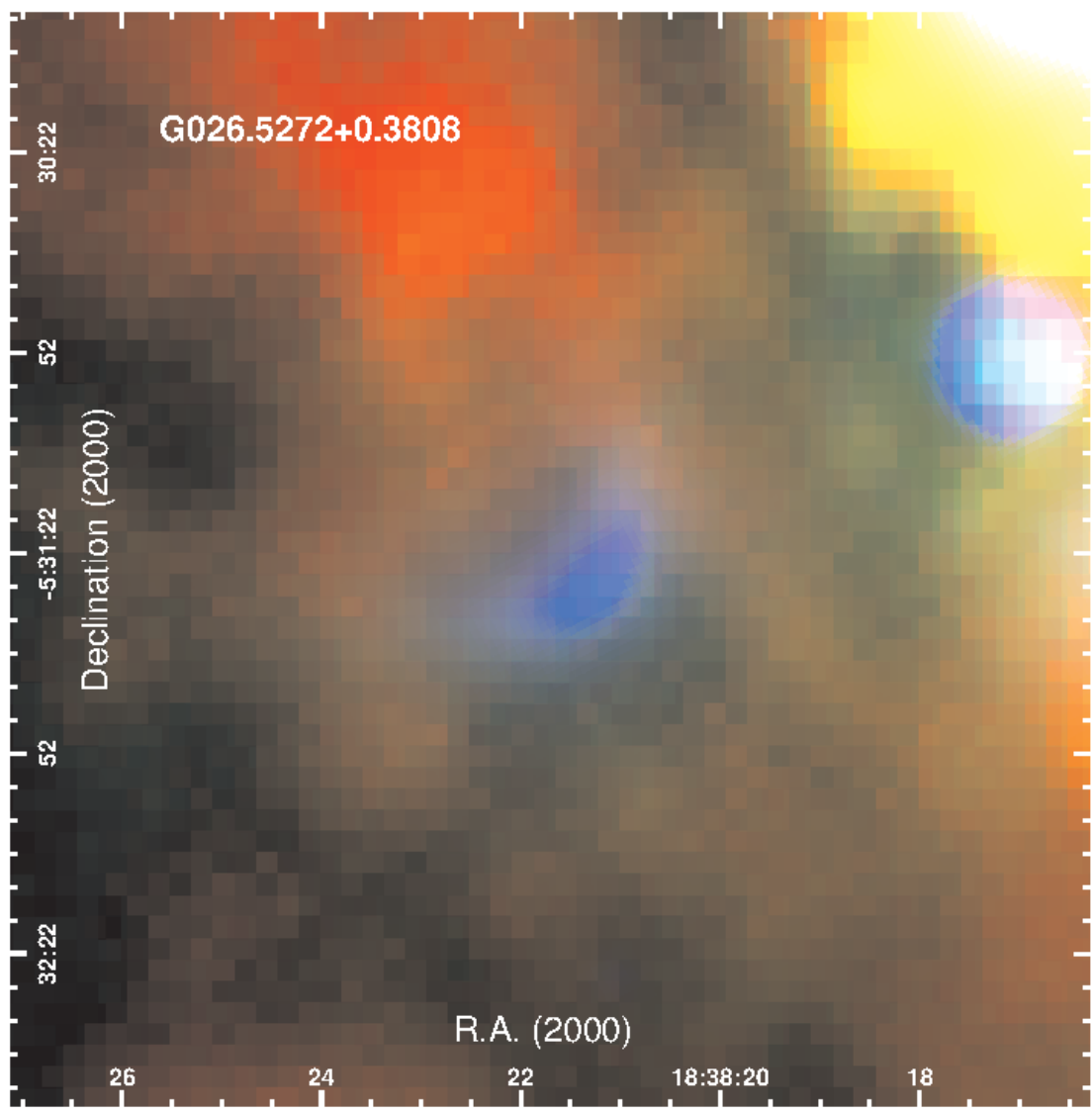}
\caption{Three-color  24/70/160 $\mu$m $SST$/$HSO$/$HSO$ depiction of the G026.5272+0.3808 nebula, as in Figure~\ref{fig:example1}. 
 This source detected at 24 and 70 $\mu$m but not
at 160 $\mu$m. \label{fig:example2}}
\end{figure}

For each object we drew a crescent-shaped polygon
encompassing the extent of the infrared nebula as seen in
the $SST$  24 $\mu$m images or $WISE$  22 $\mu$m images, as
these were the bandpasses used to first identify the
candidate bowshocks. In just a few cases, $SST$ 24 $\mu$m
data were not available  in regions nominally covered by IRAC
and MIPS.  In these cases we substituted  $WISE$ 22 $\mu$m
photometry. These exceptions are: G018.2660-00.2988,
G284.0765-00.4323, G287.4071-00.3593, G287.6131-01.1302,
G287.6736-01.0093, G288.1505.00.5059.  In seven cases
(G0150749-00.6461, G015.0812-00.6570, and G015.1032-00.6489
in the M17 region, and G284.2999-00.3359, G284.3011-00.3712,
and G284.3400-00.2827 in the Westerlund 2/RCW49 region, and
G287.4389-00.6132)  the angular size of the object was too
small and the background levels too high to carry out
meaningful photometric measurements.  Otherwise, apertures
were drawn to avoid the bright central stars which are
typically prominent at wavelengths $<$8 $\mu$m and are
occasionally detected at longer wavelengths.  We also
measured the approximate angular height, $H$, and radius,
$R$ of each nebula as an indication of its aspect ratio.
Figure~\ref{fig:example1} shows a three-color image of the
G026.1473$-$0.0420 (object \#123 from the catalog of
\citet{Kobulnicky2016}) with the source polygon outlined in
green along with  angular height and radius markers overlaid in
white.  Blue/green/red depict the 24/70/160 $\mu$m data from
$SST$/$HSO$/$HSO$,
respectively. This object shows good morphological
similarity  at all three pictured wavebands, but is most
prominent at 24 and 70 $\mu$m (blue and green).   The
complexity of the field surrounding the source and the
adjoining  bright (unrelated?) 160 $\mu$m emission to the
lower left of the source illustrates the difficulty of
determining appropriate source and background regions in
within the Galactic Plane.   Figure~\ref{fig:example2}
depicts another ``typical'' object, G026.5272+0.3808, using
the same color scheme as Figure~\ref{fig:example1} but
without the marker overlays. This source is detected at 24 and
70 $\mu$m, which show excellent morphological agreement, but
it is not detected at 160 $\mu$m.  A photometric measurement
was deemed a detection only if the morphology at a given
bandpass showed a close resemblance to that at 24 $\mu$m. 
In many cases this was a judgment call on the part of the
lead author, as the Galactic Plane exhibits a variety of
unrelated structures, especially at 160 $\mu$m, where fields
are heavily confused with foreground and/or background
emission.

Defining the shape and extent of the polygonal aperture is
unavoidably a subjective  process because the nebulae often
lie in regions of high, structured background emission that
overlap the targets.  We also defined a background polygon
for each target by displacing the source polygon to an
adjacent region judged to best reflect the local background,
typically within several arcminutes.  Our team transferred
the source and background polygons from  equatorial
J2000.0 coordinates to pixel coordinates on each of the 
relevant infrared band FITS images using {\tt
ds9}\footnote{http://ds9.si.edu/site/Home.html} image
display and analysis application.   We then performed
aperture photometry using  the {\tt IMCNTS}  task within the
Image Reduction Analysis Facility (IRAF)  {\tt
xray.xspatial} package.  The net counts (source minus
background)  were converted to Jy using fits header
information and supplementary documentation appropriate to
each of the three spacecraft observatories.\footnote{$HSO$
atlas images are in Jy pixel$^{-1}$ so no conversion after
summation is needed.  $SST$ images are in MJy sr$^{-1}$ so
the conversion factor is 10$^{6}$ Jy/MJy $\times$
206265$^{-2}$ sr/arcsec$^2$  $\times$ $S^2$
arcsec$^2$/pixel, where $S$ is the pixel size in arcseconds
appropriate to each mosaicked image, either 1.2\arcsec\ or
2.4\arcsec.  For $WISE$ the conversion from the native units
of DN to Jy is specific to each band, as given in the header
of WISE atlas images;  the conversions used here are 
1.935$\times$10$^{-6}$,
2.705$\times$10$^{-6}$,2.905$\times$10$^{-6}$,5.227$\times$10$^{-5}$ 
for bands W1, W2, W3, W4, respectively.}   The highly
variable and spatially complex  infrared structures along
the Plane, especially within the star forming regions where
some objects lie, make it challenging to define an
appropriate background region.  In the majority of cases
alternative background regions changed the photometric
measurement by 20\% or more.  Typical 1$\sigma$ photometric
uncertainties should be taken as 25\%; in some cases 50\%
would be  more appropriate.  

We did not apply aperture corrections to any of the
photometry owing to the irregular aperture shapes and sizes
which averaged 10\arcsec\ in radius but vary from 4\arcsec\
to as much as 70\arcsec. For a 10\arcsec\ radius circular
aperture, the $HSO$ PACS aperture correction is about 25\% 
\citep{Balog2014}, similar to that for $SST$ MIPS at 24
$\mu$m
\footnote{http://irsa.ipac.caltech.edu/data/SPITZER/docs/mips/mipsinstrumenthandbook/50/},
and $WISE$  at 3.4 and 4.6 $\mu$m; corrections for the
$WISE$ 12 and 22 $\mu$m bandpasses appear to be closer to
40\%
\footnote{http://wise2.ipac.caltech.edu/docs/release/allsky}.
We regard the uncertainties on the relative fluxes between
infrared bandpasses as small compared to the uncertainties
stemming from background subtraction.   Absolute fluxes
tabulated here can be regarded as 25\% low, on average, but
the fraction will vary with the size of aperture used. Use
of larger apertures was deemed inappropriate  because of the
increasing errors introduced by inclusion of unrelated
background and foreground emission.   

\begin{figure}
\plotone{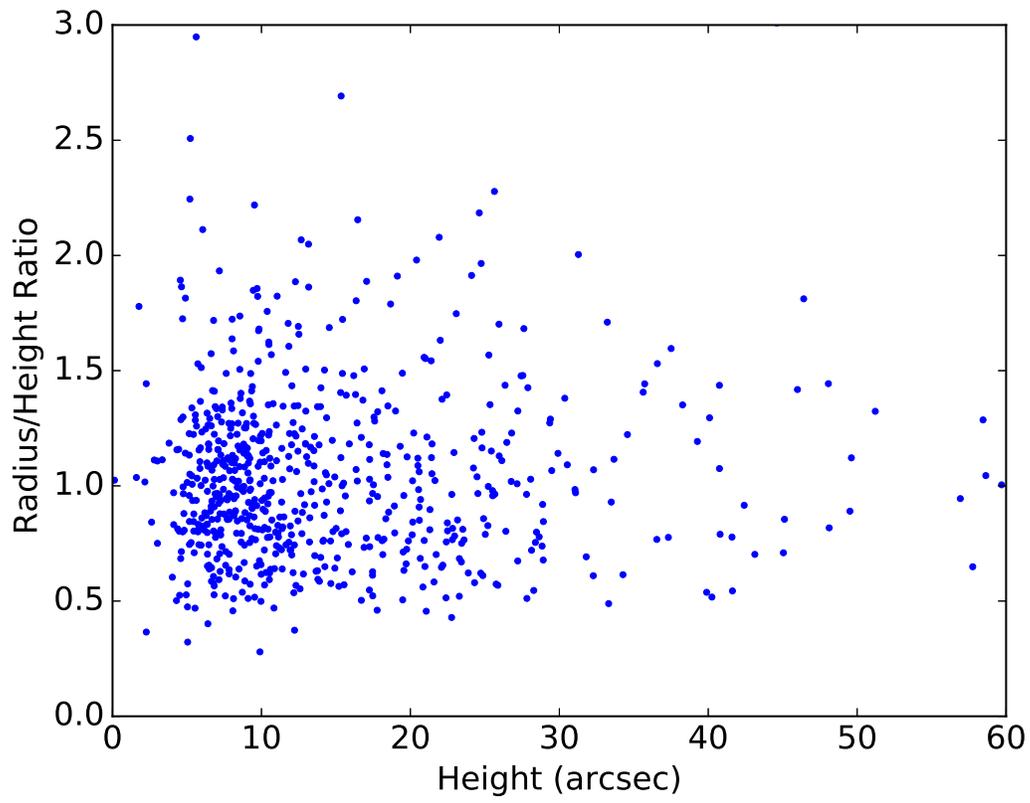}
\caption{Radius-to-height ratio versus height for arcuate nebulae illustrating the range of 
sizes and aspect ratios.  A small random offset has been added to each point to prevent overlap of 
data points at common integer values.  \label{fig:aspectratios}}
\end{figure}

Figure~\ref{fig:aspectratios} displays the ratio of nebula
radius to height (i.e., the aspect ratios) versus height.  A
small random  Gaussian offset with $\sigma=1$ arcsec 
has been added to each point to prevent overlap of  data
points at common integer values. Heights range from
3\arcsec\ in the most compact cases to over 200\arcsec\ (off
scale).   Average radius-to-height ratios near unity
indicate that most objects are approximately circular, but
values range from 0.5 to 2.0. These ratios are consistent
with the nebulae being either partial bubbles, partial
elongated bubbles, or bowshocks, which would be
morphologically indistinguishable from partial non-circular
bubbles except under the most favorable signal-to-noise
ratios and dynamic range of the image.  Additional data,
such as infrared SEDs  would allow a bowshock to be
distinguished from a bubble, given that the latter usually
display bright PAH emission in the $SST$ 8 $\mu$m or $WISE$
12 $\mu$m bandpasses where soft UV radiation from a central
star excites large molecules in a surrounding molecular
cloud \citep[e.g.,][]{Churchwell2006}.

The columns of Table~\ref{tab:Phot1} record the
identification number for each object (1)  using the
numeration of \citet{Kobulnicky2016}, its generic
designation in Galactic coordinates (2),  the Right
Ascension (3) and Declination(4) of the nominal central star
given by \citet{Kobulnicky2016}, measured radius (5) and
height (6) in arcseconds, as illustrated in
Figure~\ref{fig:example1},  fluxes in Jy in the $SST$ 3.6,
4.5, 5.8, 8.0, and 24 $\mu$m bandpasses (columns 7--11), the
fluxes in Jy in the $HSO$ 70 and 160 $\mu$m bandpasses (12,
13), and the peak 70 $\mu$m surface brightness above
background levels in Jy arcmin$^{-2}$ (14).  A value of
-99.999 indicates that no measurement was available because
either the source was not covered by the survey at this
bandpass or, in a few cases, the angular size of the source
is less than a few arcseconds and is too small for
measurement. A -99.9 in column 14 giving peak surface
brightness indicates that the source is not detected in the
$HSO$ 70 $\mu$m bandpass. Negative values for fluxes
designate $-$1 times the approximate 1$\sigma$ upper limits
for non-detections.  Column 15 gives the calculated color
temperature as derived from the 24 $\mu$m $SST$ and 70
$\mu$m $HSO$ photometric measurements.  Column 16 gives the
calculated color temperature as derived from the 70 $\mu$m 
and 160 $\mu$m $HSO$ photometric measurements.      The
first 20 lines of Table~\ref{tab:Phot1} appear in the
journal article to provide guidance as to its form and
content. The entire contents are available in the electronic
edition as a machine-readable table.  Table~\ref{tab:Phot2}
records the same information as Table~\ref{tab:Phot1} but
for the objects having four-band $WISE$ photometry instead
of five-band $SST$ photometry. The first 20 lines of
Table~\ref{tab:Phot2} appear in the journal article to
provide guidance as to its form and content. The entire
contents are online available as a machine-readable table.

Table~\ref{tab:freqs1} presents a summary of the detection
frequencies for each bandpass for the objects having $SST$
and $HSO$ photometry.\footnote{Objects detected in the $SST$
surveys are, of course, covered by the $WISE$ all-sky survey
as well, but $SST$ data are used preferentially, if
available, owing to the smaller beamsize.}  The first column
of the top row indicates that 617 objects are identified as
candidate bowshock nebulae  on the basis of $SST$ 24 $\mu$m
detections.\footnote{All but  seven of the 709
\citet{Kobulnicky2016} objects are detected at  either $SST$
24 $\mu$m or $WISE$ 22 $\mu$m. Exceptions are \#51, \#52,
\#53 --- in the M17 star-forming region --- \#400, \$401,
\#402 --- in the RCW~49 star-forming region ---, and \#408
---  in the Carina star-forming region --- which are
identified on the  basis of $SST$  8.0 $\mu$m or shorter
wavelength images.}   The second column the top row
indicates that 399 of these also have detections at the
$HSO$ 70 $\mu$m bandpass.   The last column of the top row
indicates that only 35 of these 339 are further detected in the
$HSO$ 160 $\mu$m bandpass.  A detection at 160 $\mu$m always
implies a detection at 70 $\mu$m.   The second row reports
detections at the $SST$ 8.0 $\mu$m bandpass in conjunction
with detections at longer wavelengths.   The 106 in the
first column gives the number of objects detected at both 24
$\mu$m and 8.0 $\mu$m.   The 90 in the second column gives
the number that are further detected at 70 $\mu$m.   The
final column shows that only 20 objects are detected at 8.0
$\mu$m and all three longer wavelengths.  The bottom row of
Table~\ref{tab:freqs1} reports the number of objects
detected at 3.6 $\mu$m and successively longer wavelengths. 
Thirty-two objects are detected at 3.6 $\mu$m  and at all
longer $SST$ wavelengths.  Only 26 have detections at 3.6
$\mu$m through 70 $\mu$m, and only eight are detected at all
bandpasses from 3.6 $\mu$m through 160 $\mu$m.    

Table~\ref{tab:freqs2} reports detection frequencies for
objects measured in the $WISE$ and $HSO$ survey images. The
first column of the top row (85) designates the number of
catalog objects detected exclusively in the $WISE$ 22 $\mu$m
data.  The second and third  columns give the number of 
objects further detected at 70 $\mu$m and 160 $\mu$m (23 and
4, respectively).  The middle row lists the number of
objects detected at the $WISE$ 12 $\mu$m bandpass, and
successively longer wavelengths. A detection at 22 $\mu$m is
almost always accompanied by a detection in the 12 $\mu$m
bandpass which includes both PAH and hot dust
contributions.  The bottom row gives the number of
detections at  the $WISE$ 3.4 $\mu$m bandpass along longer. 
Only 12 objects are detected at 3.4 $\mu$m through 22
$\mu$m;  two of these have 70 $\mu$m detections, and only
one has a 160 $\mu$m detection.

\section{Infrared Color Analysis}

Photometric measurements at two or three bandpasses ---
available for the majority of this sample --- enable 
calculation of the infrared colors and  color temperatures.
We undertake this type of analysis first in Section 3.1 as a
means of characterizing the ensemble properties of a large
number of bowshock nebula candidates.  For objects where the
temperature, radius, and distance of the central 
illuminating star is known, more sophisticated types of
analyses are possible.  These include comparisons of the
expected steady-state  dust temperatures and radiant energy
densities in the nebulae to the temperatures and energy
densities implied by fitting  interstellar dust models to
the SEDs.  We evaluate these additional diagnostics in
Sections 3.2 and 3.3 for a subsample of 20 nebulae
having well-characterized central stars.  

\subsection{IR Colors}

\begin{figure}
\plotone{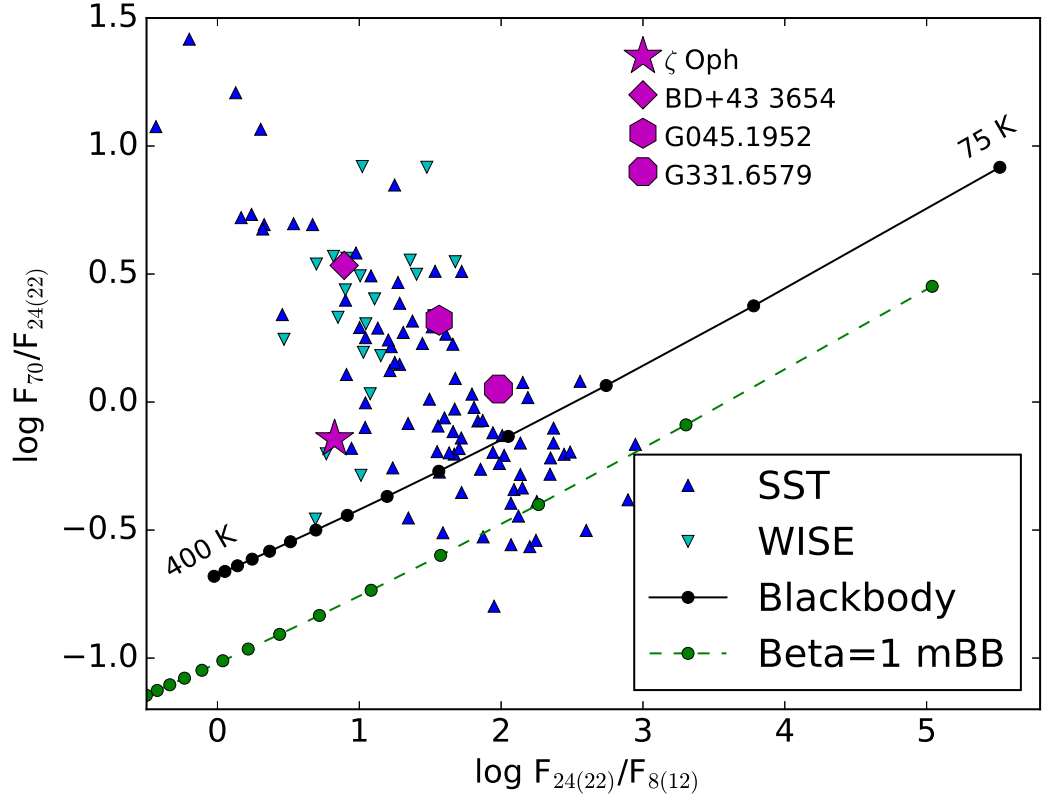}
\caption{Color-color plot of candidate bowshock nebula photometry showing the ratio of flux at $HSO$ 70 $\mu$m / $SST$ 24 $\mu$m versus ratio of flux at $SST$ 24 $\mu$m / $SST$ 8 $\mu$m ({\it upward pointing triangles)}.  Objects detected with $WISE$ have photomery at 22 $\mu$m instead of 24 $\mu$m and 12 $\mu$m instead of 8 $\mu$m ({\it downward pointing triangles)}.  The solid curve illustrates the colors of a blackbody denoted by filled circles every 25 K from 75 K at upper right to 400 K at lower left.  The dashed curve 
illustrates the colors of a modified blackbody with $\beta=1$ over the same range.  A star and diamond denote the colors of the bowshock nebulae associated with the well-established runaway stars $\zeta$ Oph and BD+43~3654.
A hexagon and octogon denote colors of G045.1952+00.7420 and G331.6579+00.1308. \label{fig:color1}}
\end{figure}

Ratios of infrared fluxes provide constraints on the
properties of the emitting material, including dust
temperatures.  Figure~\ref{fig:color1} shows a color-color
diagram for the 115 nebulae detected at  the $HSO$ 70 $\mu$m
band and $SST$ 24 and 8 $\mu$m ({\it upward pointing
triangles}; 92 objects) or $WISE$ 22 and 12 $\mu$m ({\it
downward pointing triangles}; 23 objects).  $SST$ and $WISE$
objects exhibit a large degree of overlap, together forming
a band that stretches from upper left (high 70/24 $\mu$m and
low 24/8 $\mu$m ratios) to lower right (low 70/24 $\mu$m and
high 24/8 $\mu$m ratios). The $WISE$ targets lie
preferentially in the upper half of the $SST$ objects.   The
solid curve illustrates the colors of a blackbody denoted
by  filled circles every 25~K from 75~K at upper right to
400 K at lower left. The dashed curve illustrates the colors
of a modified blackbody ($S_\nu \propto \nu^\beta B_\nu(T)$)
with $\beta=1$ over the same range.  That the distribution
of points runs orthogonal to the blackbody curves is an
indication that the objects are not well characterized as
single-temperature dust in the majority of cases.  If
anything, the modified blackbody shows poorer agreement with
the data than the simple blackbody.  A star and diamond
denote the colors of the canonical bowshock nebulae
associated with the runaway stars $\zeta$ Oph and
BD+43~3654. A hexagon and octagon denote colors of
G045.1952+00.7420 and G331.6579+00.1308, as indicative of
other bright objects in our sample.  Typical photometric
uncertainties are about 0.2 dex on each axis.  Most of the
points lie to the upper left of the blackbody curve,
indicating that their color temperatures inferred from the
$F_{70}$/$F_{24}$ ratios are considerably cooler than those
indicated by the $F_{24}$/$F_{8}$ (or $F_{22}$/$F_{12}$)
ratios.  This discrepancy may result from a  contribution
from very large molecules (i.e., PAH's) or an anomalously
large population of small, hot, non-thermally heated grains
emitting in the $SST$ IRAC 8 $\mu$m and $WISE$ 12 $\mu$m
bandpasses. Such an effect would  lower the $F_{24}$/$F_{8}$
ratio and shift points horizontally to the left of the
blackbody curve.  For this reason, the $F_{70}$/$F_{24}$
ratio is likely to be a more reliable  indicator of the dust
temperature.  However, most points lie several orders of
magnitude to the left of the blackbody fiducial, making it
unlikely that excess contribution from PAHs is (solely)  the
cause.  Emission from dust at multiple temperatures within
an extended dust and gas nebula likely plays a role.

\begin{figure}
\plotone{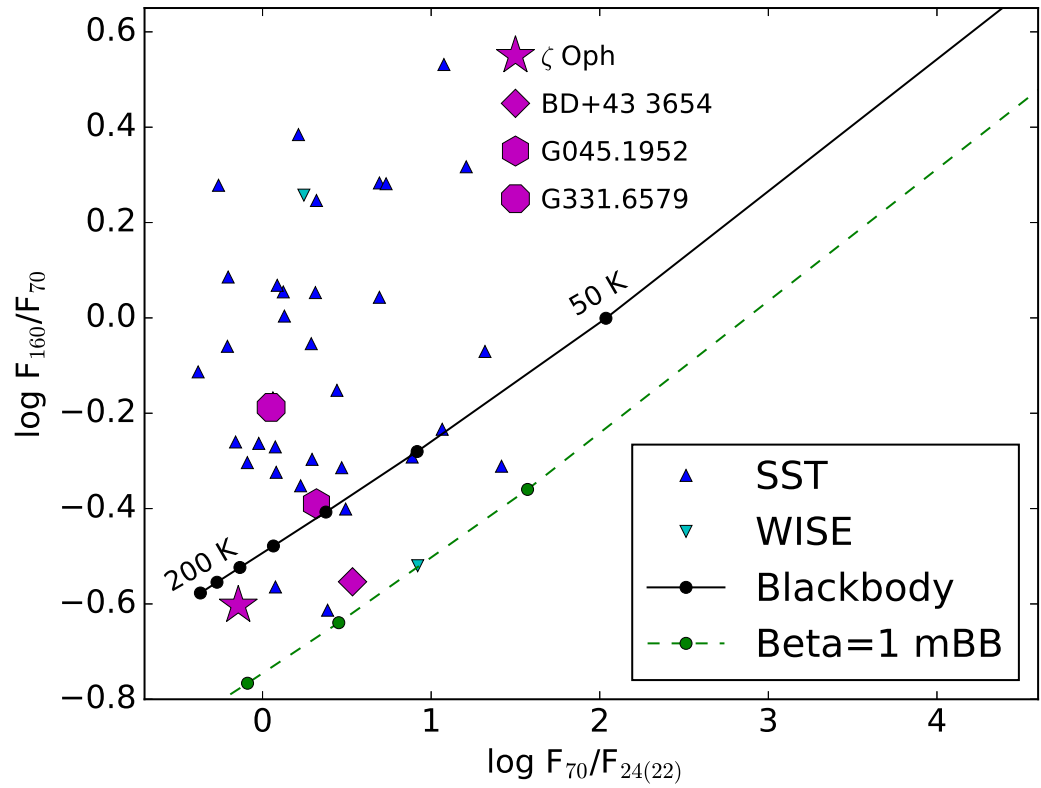}
\caption{Infrared color-color diagram for nebulae detected at the $HSO$ 160 and 70 $\mu$m bandpasses and at the $SST$ 24 $\mu$m or $WISE$ 22 $\mu$m bandpass.  Symbols are the same as in Figure~\ref{fig:color1}.    \label{fig:color2}}
\end{figure}

Figure~\ref{fig:color2}  plots the log of the
$F_{160}$/$F_{70}$ ratios versus log of the
$F_{70}$/$F_{24}$ ratios for the 36 $SST$ objects and 4
$WISE$ objects having data.  Symbols are the same as in
Figure~\ref{fig:color1}.  As in  Figure~\ref{fig:color1} the
majority of points lie to the upper left of the blackbody
curve.  In these bandpasses PAHs are not present, so another
cause is required to explain the distribution in this color
space.  One possible explanation is, again, an excess of
very small non-thermally heated grains that contributes
disproportionatly at the 24/22 $\mu$m bands, shifting points
to the left of the blackbody curve.   Another possibility is
a multi-temperature dust structure.  The apparent excess of
$F_{70}$ relative to $F_{24}$ in Figure~\ref{fig:color1}
and  excess of $F_{160}$ relative to $F_{70}$ in
Figure~\ref{fig:color2} would be consistent with the
majority of sources showing increased contributions from
cool dust at longer wavelengths.  Infrared morphologies
provides some support for this interpretation.

\begin{figure}
\gridline{\fig{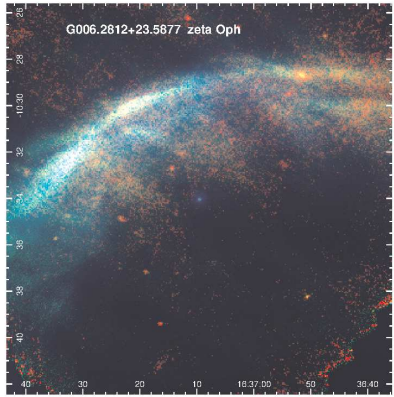}{0.3\textwidth}{(a)}
	  \fig{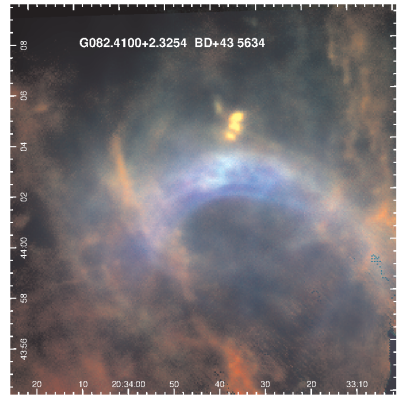}{0.3\textwidth}{(b)}}
\gridline{\fig{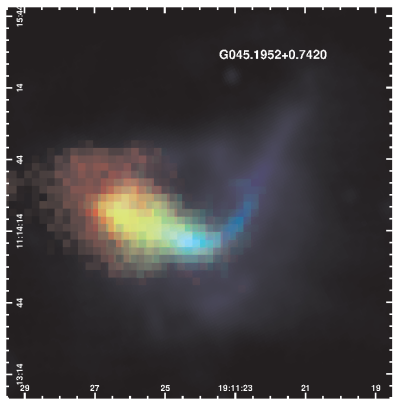}{0.3\textwidth}{(d)}
	  \fig{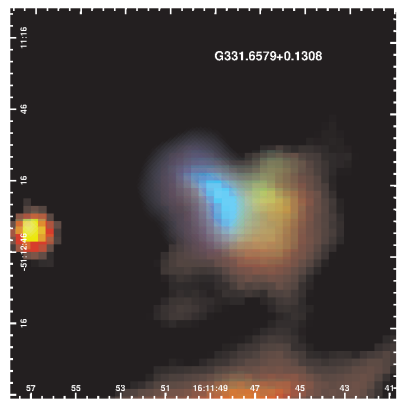}{0.3\textwidth}{(f)}}
\caption{Three-color 160/70/24 (or 22) $\mu$m images in red/green/blue, respectively, of four nebulae:
the canonical runaway star and bowshock nebula $\zeta$ Oph (upper left), BD+43~3654 (upper right), G045.1952+00.7420 (lower left), and  
G331.6579+00.1308 (lower right). \label{fig:4examples} }
\end{figure}

Figure~\ref{fig:4examples} displays four examples of the
160/70/24(22) $\mu$m morphologies in red/green/blue,
respectively, for the four objects plotted with magenta
symbols in the preceding color-color diagrams.  The upper
left panel is the bowshock associated with the O9.2IV star
$\zeta$ Oph (object \#13).  It spans more than 10\arcmin\ on
the sky but has high quality measurements  from $HSO$ and
$WISE$ data owing to its high Galactic latitude (+23\degr)
and, therefore, relative lack of foreground/background
contamination.  With a HIPPARCOS parallax of 8.91 mas its
implied distance of 112 pc makes $\zeta$ Oph the nearest
known bowshock. Consequently, its nebular appearance is
highly resolved into wispy filaments better seen in
mid-infrared $WISE$ bandpasses.\footnote{E.g., 
https://www.nasa.gov/mission$\_$pages/WISE/news/wise20110124.html.} 
This object, plotted as a magenta star, lies near the
blackbody curve in Figures~\ref{fig:color1} and
\ref{fig:color2}.   Somewhat smaller in angular size is the
nebulae associated with the O4If star BD+43~3654 (upper
right; object \#344).  This object is prominent at 22 $\mu$m
(blue) but less distinct at the $HSO$ bandpasses.  At
Galactic latitude +2.3\degr, it also suffers from less
background confusion than a typical Galactic Plane source,
but it is located  near the extended Cygnus-X star-forming
complex at a probable distance of 1.3--1.5 kpc
\citep{Kiminki2015, Rygl2012} which contributes to unrelated
line-of-sight emission.  This object, plotted as a diamond,
lies far from  the blackbody curve in
Figure~\ref{fig:color1} but near the modified blackbody
curve in Figure~\ref{fig:color2}. G045.1952+0.7420 (lower
left; object \#238) illustrates the not-uncommon
circumstance where the 24 $\mu$m arcuate nebula has 70 and
160 $\mu$m counterparts showing  significantly different
morphologies.  Azimuthal variations in dust temperature and
density seem likely in objects such as this.  This object,
plotted as a hexagon, lies far from  the blackbody curve in
Figure~\ref{fig:color1} but very near the blackbody curve in
Figure~\ref{fig:color2}. G331.6579+0.1308 (lower right;
object \#582) exemplifies another pattern noticed in about
10--15\% of objects.  The prominent 24 $\mu$m arc is
accompanied by 70 and 160 $\mu$m features that are offset
spatially to the exterior of the arc.  This object, plotted
as a hexagon, lies near the blackbody curve in
Figure~\ref{fig:color1} and somewhat further from the
blackbody curve in Figure~\ref{fig:color2}.

\begin{figure}
\plotone{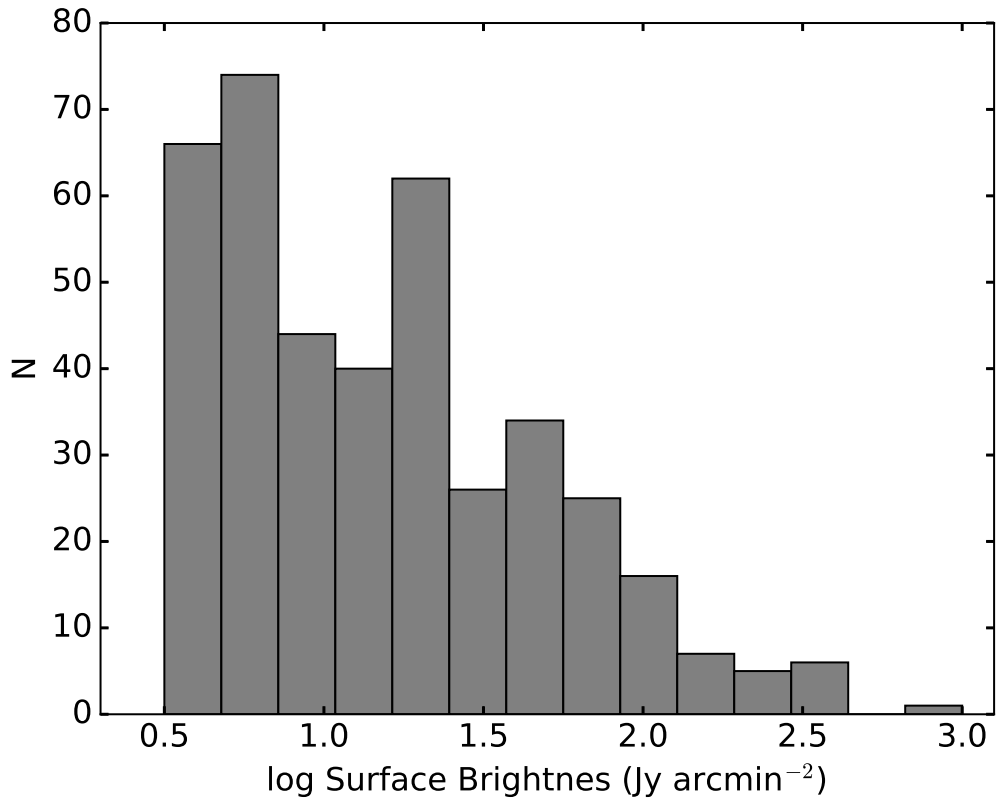}
\caption{Histogram of peak 70 $\mu$m surface brightness for target objects.  \label{fig:surface}}
\end{figure}

The  morphology in the lower right panel of
Figure~\ref{fig:4examples} might result if a stellar
wind-driven bowshock is encountering a molecular cloud or
region of higher density.  Such an arrangement provides a
possible explanation of why only 10\% of runaway stars
produce detectable infrared bowshocks \citep{Peri2015}. 
Perhaps a sufficiently high dust density or a density
gradient is required to produce the requisite surface
brightness to be detected amidst the high background  levels
in the Plane.  For example, if $\zeta$ Oph, with a peak 70
$\mu$m surface brightness of 9 Jy arcmin$^{-2}$ were placed
in the Plane where our typical objects have surface
brightnesses of tens to hundreds of Jy arcmin$^{-2}$, it
would likely go unnoticed and be undetectable. At 22 $\mu$m
the $\zeta$ Oph bowshock  nebula has a surface brightness 
of 1.9 Jy arcmin$^{-2}$, which may be considered typical of
our sample where  values range from 0.6 to over 20 Jy
arcmin$^{-2}$.   Figure~\ref{fig:surface} shows a histogram
of 70 $\mu$m surface brightness for our sample.  Values
range from 3--300 Jy arcmin$^{-2}$. The limit of
detectability is usually around 0.02 Jy pixel$^{-1}$  or 7
Jy arcmin$^{-2}$ in the $HSO$ 70 $\mu$m images, but in some
favorable sightlines factors of two lower are possible.  For
some objects that are just marginally detected, a value of
3.5  Jy arcmin$^{-2}$ is a typical value, representing a
rough lower limit to detectability.    

\subsection{Dust Temperatures}
   
\begin{figure}
\plottwo{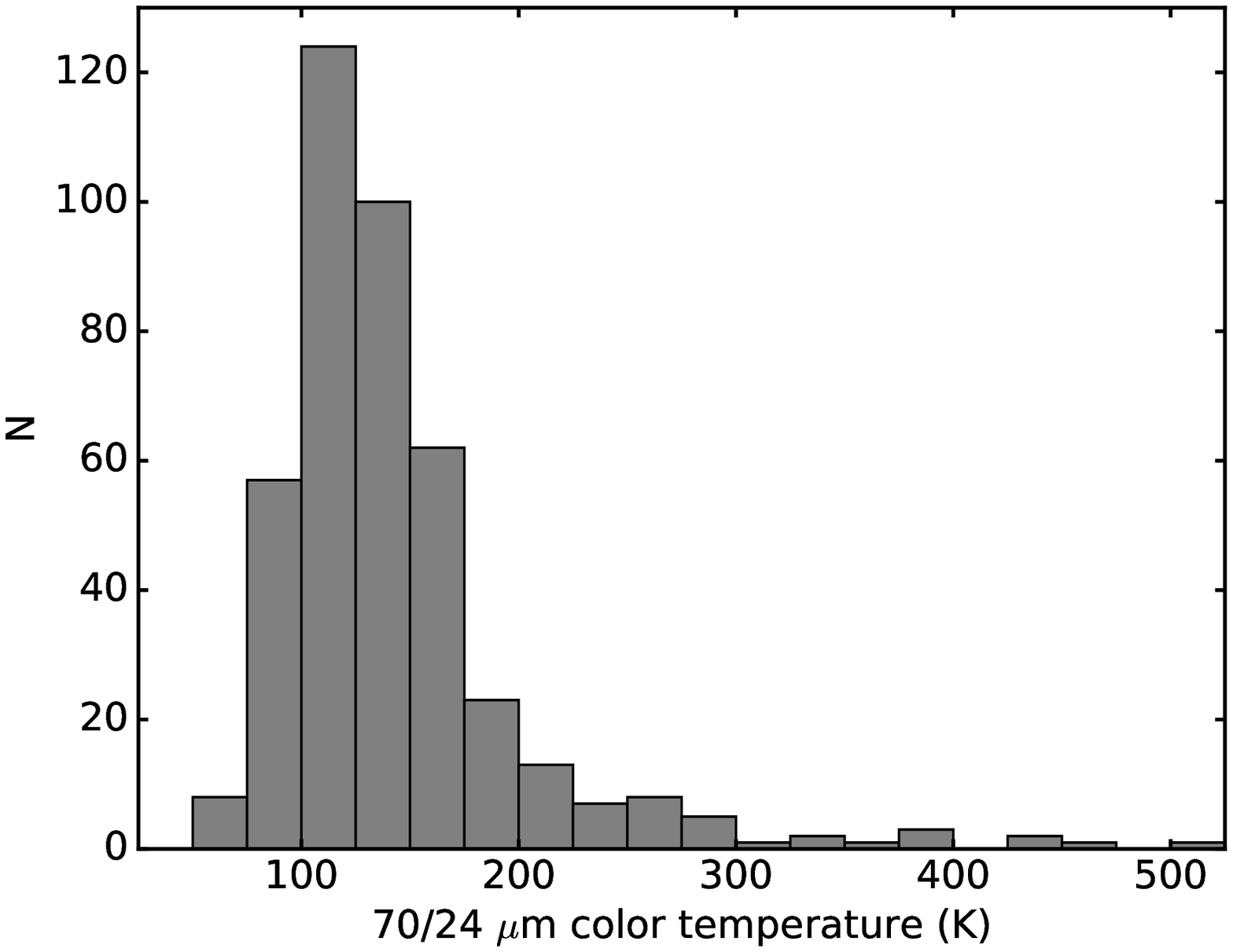}{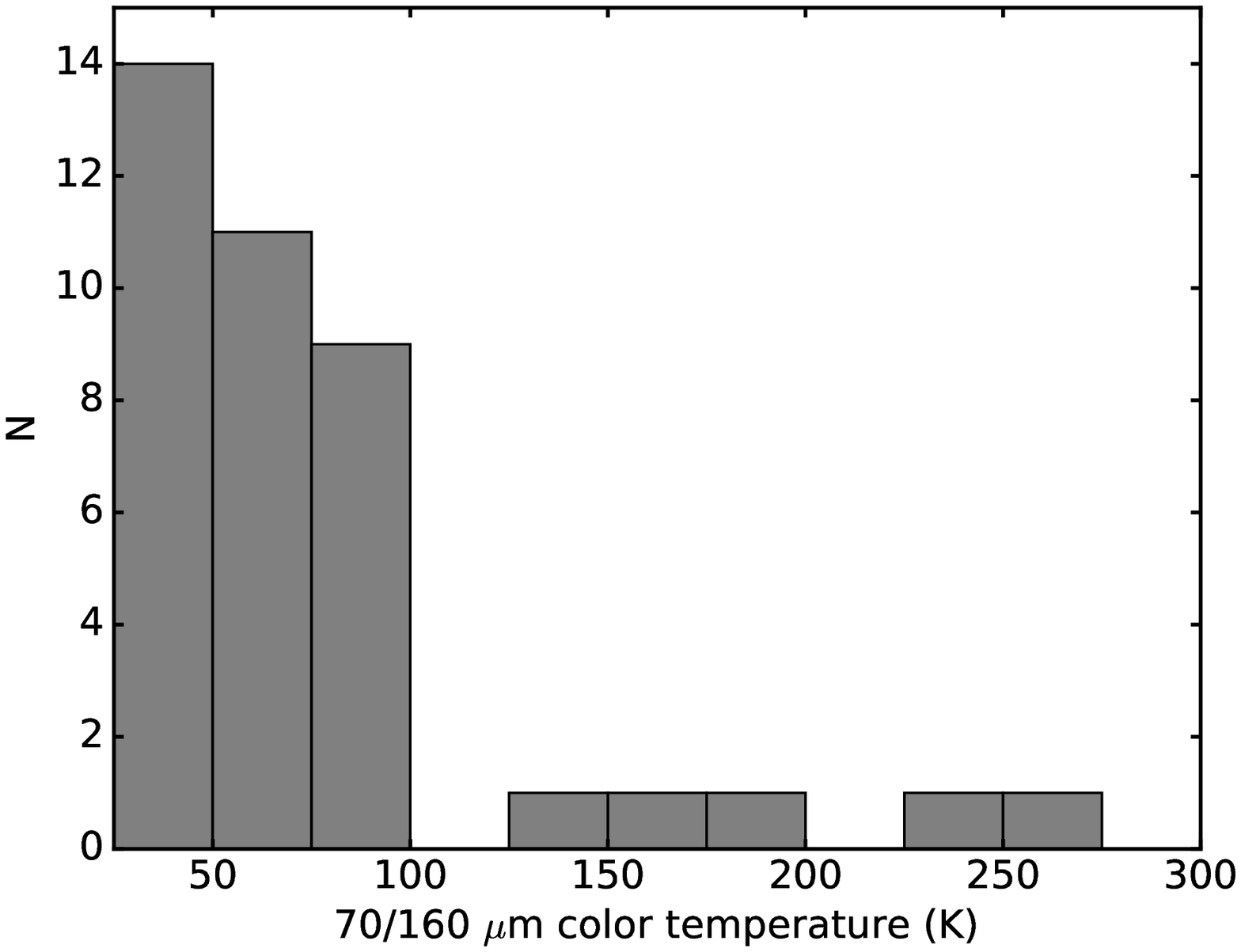}
\caption{Histogram of color temperatures determined from 24 and 70 $\mu$m measurements (left) and
histogram of color temperatures determined from 70 and 160 $\mu$m measurements (right). 
The long tail toward higher dust temperatures in the left panel is plausible the result of 
an overabundance of small stochastically heated grains emitting at 24 $\mu$m.  
\label{fig:temps}}
\end{figure}

\begin{figure}
\plotone{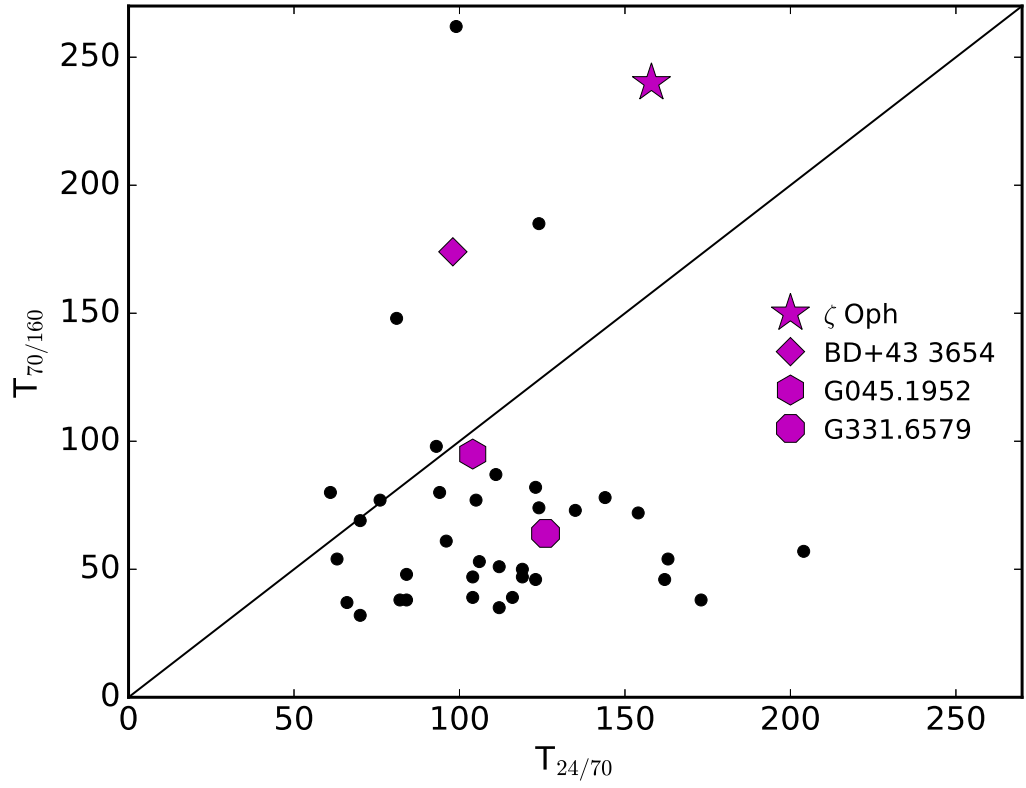}
\caption{Color temperatures derived from 70 and 160 $\mu$m measurements versus those from 24 and 70 $\mu$m measurements. Filled large symbols are the same objects as in Figure~\ref{fig:color1}. \label{fig:comparetemps}}
\end{figure}

Figure~\ref{fig:temps} (left panel) shows a histogram of
color temperatures calculated by fitting a blackbody to the
24 (or 22) and 70 $\mu$m data for the 422 objects having 
measurements at those two bandpasses.   The distribution is
approximately Gaussian with a mean of 147 K, an rms of 78 K,
and a long tail toward higher temperatures as large as
500~K.  Figure~\ref{fig:temps} (right panel) shows a
histogram of color temperatures for the 40 objects having
both 70 and 160 $\mu$m $HSO$ measurements.  Nearly all the
objects cluster between 25 and 100 K, with a few outliers up
to 300~K.   Figure~\ref{fig:comparetemps} plots the 70
$\mu$m/160 $\mu$m color temperatures versus the 24 $\mu$m/70
$\mu$m color temperatures.  The vast majority of objects lie
to the right of the 1:1 relation, and there is no
significant correlation.  We interpret this to mean either
that the 160 $\mu$m measurements are  unreliable, sampling
unrelated foreground or background material, or that the
arcuate nebulae are intrinsically multi-temperature
structures, as might be expected in the presence of a
density gradient such as at the edge of a molecular cloud. 
This discrepancy could also result from a population of
small, hot, semi-stochastically heated grains elevating the
24  $\mu$m fluxes and skewing the fitted color temperature
upward toward higher values; such a scenario would be
indistinguishable from the multi-temperature hypothesis on
the basis of only three broadband measurements.   Images
presented above show that the 160 $\mu$m emission is
sometimes offset from the 24 $\mu$m emission, lending
credence to the idea of multiple temperature components. 
Shocks may also fragment dust grains producing a population
of stochasitically heated very small grains radiating
strongly in the 24 $\mu$m band.

The nebula associated with $\zeta$ Oph (plotted as a magenta
star) has one of the highest 70/160 $\mu$m color
temperatures in the sample.  Its 70/160 $\mu$m color
temperature is 240 K versus a 24/70 $ mu$m temperature of
158 K.  We believe this is largely the result of the large
angular size of the source (over 400\arcsec\ in diameter)
which makes the photometry, particularly at 160 $\mu$m, very
sensitive to the adopted background level.  The  BD+43~3654
nebula (magenta diamond) also lies above the 1:1 line,
plausibly for similar reasons. Although it is considerably
smaller at an angular diameter of about 120\arcsec, it lies
close to the Plane in a confused region near the Cygnus OB2
star-forming complex. The G045.1952+00.7420 nebula (hexagon)
has very similar color temperatures from both diagnostics. 
The G331.6579+00.1308 nebula lies below the line along with
the majority of the objects.  It has a 24/70 $\mu$m
temperature of 126 K and 70/160 $\mu$m temperature of 64 K. 
As illustrated by the lower right panel of
Figure~\ref{fig:4examples}, this object has an extended area
of 160 $\ mu$m emission just outside the arcuate region
defined by 24 $\mu$m emission.  Inclusion of this, perhaps
unrelated material, would lower the derived color
temperature.  The majority of objects in
Figure~\ref{fig:comparetemps} probably are affected by a
similar circumstance.  We consider the 70/160 color
temperatures to be less reliable and/or less diagnostic than
the temperatures 24/70 color temperatures, even allowing for
the possibility of small stochastically heated grains
affecting the 24 $\mu$m bandpass.  

\section{SED modeling}

Detailed modeling of the spectral energy distributions  is
possible for a few objects where the properties of the
central stars ($T_{eff}$, $R_*$) are known and distances to
the objects are also well constrained so that the standoff
distance, $R_0$, between the star and nebula is also known.
The star must also have at least two, and preferably three
or more photometric measurements.   
Table~\ref{tab:Tcompare} lists 20 objects that meet these
criteria.  Columns 1--3 give the target ID number used in
this work and in \cite{Kobulnicky2016}  along with the
common name and identifier in Galactic coordinates.   Column
4 gives the spectral type and luminosity class as listed in
the literature or, in a few cases, from our own spectroscopy
\citep{Chick2017}.  Column 5 is the adopted effective
temperature, $T_{eff}$,  from the assigned spectral type and
luminosity class using the theoretical effective temperature
scale of \citet{Martins2005}. Column 6 is the corresponding
stellar radius, $R_*$, in solar radii.  Column 7 lists the
adopted distance, $D$, to each source in kpc.   For the
majority of the objects in Table~\ref{tab:Tcompare} 
distances are estimated through their association with a
star cluster or molecular cloud of known distance, measured
from main sequence color-magnitude diagram fitting, or from
radio very long baseline interferometry parallax
measurements toward masers.\footnote{Spectrophotometric
distance estimates toward the remainder of our sample are
possible but considerably more uncertain owing to
difficulties in assessing accurate spectral types and
luminosity classes,  reddening and binarity.  We defer this
analysis for a future work.}     Instances include NGC~6611
at 1.99 kpc \citep{Hillenbrand1993}, the W3/4/5 star forming
complex at 2.00 kpc \citep{Xu2006}, the Cygnus X
complex/Cygnus OB2 at 1.32 kpc \citep{Kiminki2015,Rygl2012},
and the Carina star forming complex at 2.3 kpc
\citep{Allen1993}. The remainder are sufficiently close to
have optical parallax measurements
\citep[HIPPARCOS;][]{Perryman1997}.  Columns 8--9 list the
standoff distance of the nebulae, $R_0$, from the central
star in arcsec and pc, respectively.  From these data we
compute, in Column 10, the radiation density parameter, $U$,
at the location of the nebula. $U$ is defined in
\citet[][DL07]{DL07}  as the ratio of the radiant energy
density (in erg~cm$^{-3}$) due to the star to the mean
interstellar radiant energy density estimated by
\citet[][MMP83]{Mathis1983},

\begin{equation}
U = {u_* \over u_{MMP83}} = { {R_*^2 \sigma ~T_{eff}^4 / (R_o^2 c)}  \over {0.0217 erg~s^{-1} cm^{-2}}/c }. 
\end{equation}

\noindent \citet{DL07} use this dimensionless ratio, $U$, 
as a parameter characterizing the radiant energy density in
their grid of model  dust emissivities.  For the objects in
Table~\ref{tab:Tcompare} these values range from several
$\times$10$^2$ to 2$\times$10$^5$, indicating that the
central stars dominate the radiant energy density  at the
locations of the nebulae.  From these basic data it is
straightforward to compute the steady state  temperature of
dust, $T_{SS}$, in the nebulae if radiant heating from the
central star were the dominant heat source.  Adopting 
\citet{Draine2011} equation 24.19  for the temperture
approximation of siliate grains in the size range $0.01<a<1$
$\mu$m,

\begin{equation}
T_{SS}~(K) = 16.4 ~ (a/0.1 \mu m)^{-1/15} U^{1/6}, 
\end{equation}

\noindent(where $a$ is the grain size, taken to be
0.1 $\mu$m) for silicate dust.\footnote{$T_{SS}$, listed in
Column 11,  becomes $\simeq$35\% larger for graphite dust 
\citep[equation 24.20]{Draine2011}. }   $T_{SS}$ is
relatively insensitive to errors in $R_*$ or $R_0$ (derived
from the adopted distance), given that $T_{SS}$ scales
weakly  with both quantities.  $R_0$ is actually a lower
limit on the star-nebula separation since the inclination
angle of the vector from star to nebulae apex is unknown.  
Nevertheless, the inclination angle is likely to be near
90\degr\ owing to selection biases that would work against
detection of bowshocks viewed at small inclination angles. 
Furthermore, most dust within the nebula lies at distances
$>R_0$  from the star, given the approximately parabolic
bowshock shape,  making the derived T$_{SS}$ a firm upper
limit.  

\begin{figure}
\plotone{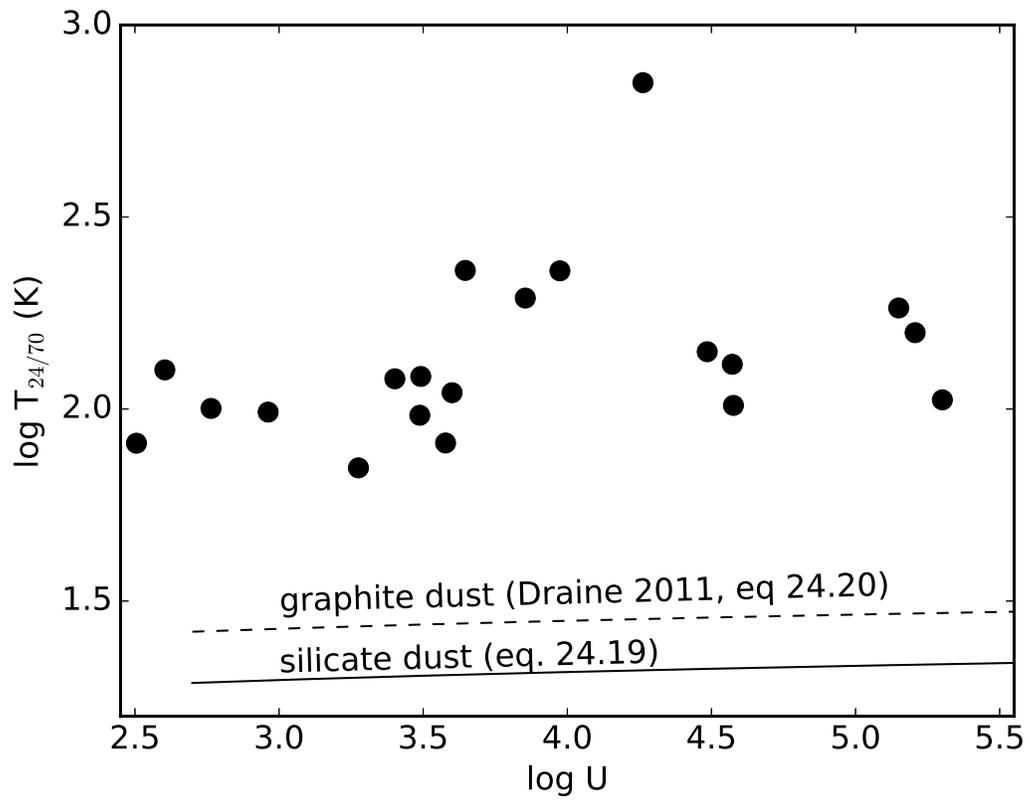}
\caption{Color temperatures derived from  24 and 70 $\mu$m measurements
versus radiation energy density parameter, $U$. 
The solid and dashed line show the predicted relationships
from \citet{Draine2011} for silicate and graphite dust, respectively. \label{fig:TvsU}}
\end{figure}

Columns 12 and 13 of Table~\ref{tab:Tcompare} are the color
temperatures derived from  the 24/70 or 70/160 $\mu$m
ratios.   Figure~\ref{fig:TvsU} plots the log of derived
24/70 $\mu$m  color temperatures versus log $U$.  The solid
line shows the relation predicted by \citet{Draine2011},
equation 24.19 appropriate to silicate dust while the 
dashed line shows the prediction for graphite dust. It is
immediately apparent that the predicted steady state dust
temperatures are  smaller than the measured color
temperatures by factors of 1.1--3, with a mean ratio of
1.76.   We interpret  this as evidence that the infrared
nebulae are not merely interstellar  material sculpted and
heated radiantly by the central star.    Evidently, these
nebulae require an additional source of heating, such as
shock heating from the stellar wind mechanical energy.  The
discrepancy between the 24/70 $\mu$m color temperatures and
steady state dust temperatures could also be explained by
invoking a population of anomalously small grains with $a
\ll 0.1~{\mu}m$, but given the $a^{-1/15}$ dependency,
increasing the dust temperature by 30\% would require a
large deviation from the interstellar grain size
distribution adopted in \cite{Draine2011}.  Such a
preponderance of small grains, if confirmed,  would support
the idea of grain fragmentation in the stellar wind bow
shocks.  Mid-infrared spectroscopy of the nebula  in
conjunction with dust models including excess populations of
small grains could potentially allow confirmation of a
variance from the interstellar grain size distribution. 

Column 14 of Table~\ref{tab:Tcompare} contains the total
flux, of the nebula ($R_{IR}$, in  erg s$^{-1}$ cm$^{-2}$) 
estimated by integrating a blackbody fit to the 24 (or 22) 
and 70 $\mu$m data points over the range 1 $\mu$m to 300
$\mu$m. As demonstrated  below, a blackbody provides a
reasonable fit to the overall SED at wavelengths longer than
22 $\mu$m in most cases.  Column  15 contains the ratio of
stellar luminosity to nebular luminosity, $L_*/L_{IR}$. 
$L_*$ is calculated using the adopted $R_*$ and $T_{eff}$.  
Luminosity ratios range from 8 to 5100, with typical values
of several hundred.   Apparently the  dust in the nebulae
intercepts and re-radiates a small faction of the stellar 
luminosity.  There is no correlation between $L_*/L_{IR}$
and the standoff distance, $R_0$, as might be  expected if
larger standoff distances result in the nebulae intercepting
and re-radiating a smaller fraction of the star's
luminosity.

\subsection{SEDs of Sources with 160 $\mu$m Detections}

\begin{figure}
\plotone{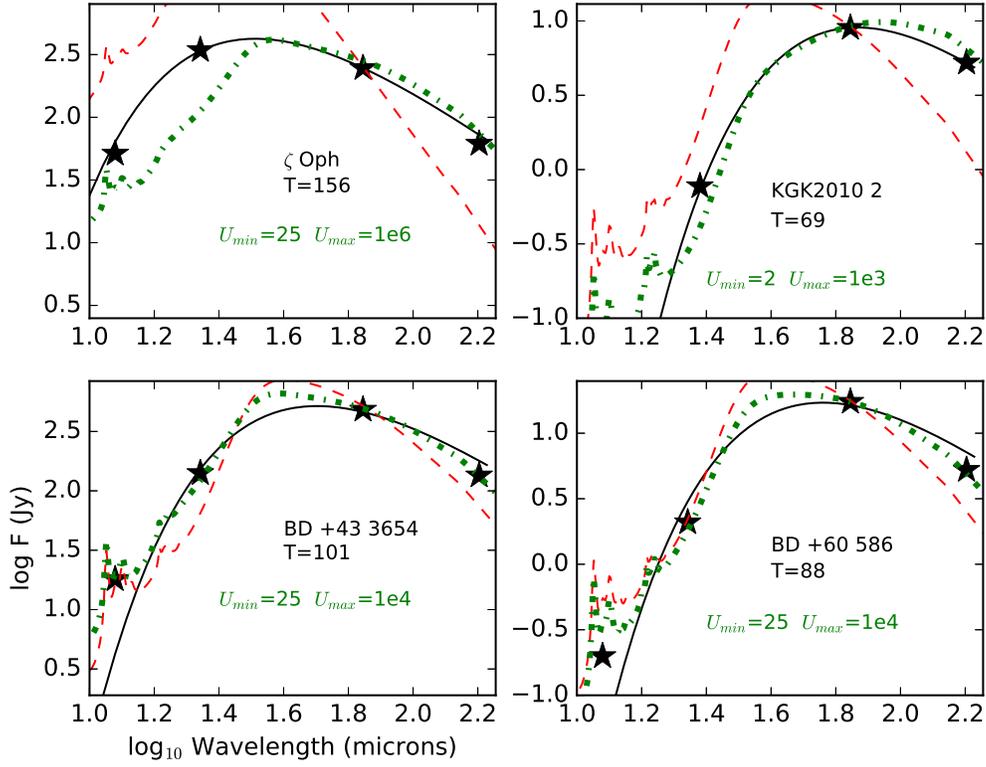}
\caption{Spectral energy distributions for four nebulae having well-characterized central stars and distances from Table~\ref{tab:Tcompare}. 
Points show photometric data at $WISE$, $SST$, and $HSO$ bandpasses. Black solid curves are single-temperature blackbody fits.
Dashed red curves are DL07 dust models for fixed $U$, the scaling
parameter for the radiation field intensity, appropriate to the central star's luminosity and the nebula's standoff distance, $R_0$, as listed in 
Table~\ref{tab:Tcompare}.
Dash-dot green curves are DL07 dust models having a range of radiation
energy density between $U_{min}$ and $U_{max}$. All models have the minimum PAH contribution of 0.47\% by mass.  
\label{fig:SED1}}
\end{figure}

Knowledge of the radiation field can be used in conjunction
with the  \citet{DL07} interstellar dust models to infer the
properties of dust in the subsample of well-characterized
bowshock nebula candidates.   The grid of \citet{DL07}
is parameterized in terms three variables: dust exposed to a
minimum radiant energy density $U_{min}$, a maximum radiant
energy density $U_{max}$, and a fraction of PAH molecules by
mass $q_{PAH}$. We use the models appropriate to Milky Way
dust and metallicity.\footnote{\citet{DL07} also consider
linear combinations of dust heated by a single $U$ and dust
heated by a range of $U$, via their parameter, $\gamma$,
sspecifying the fraction of dust mass exposed to each
radiant energy density.  Given the limited number of
photometric data points available in our study, we do not
include this additional free parameter in our model
comparisons.}   Figure~\ref{fig:SED1} plots the infrared
spectral energy distributions for the four objects from 
Table~\ref{tab:Tcompare} that have 160 $\mu$m $HSO$
measurements. Points show photometric data at $WISE$ 12 and
22 $\mu$m, $SST$ 24 $\mu$m, and $HSO$ 70 and 160 $\mu$m
bandpasses.  Typical photometric uncertainties are 0.2 dex. 
Black solid curves are best-fitting blackbodies, with labels
in each panel indicating the temperature.  The blackbody
curves fit the data well in the upper two panels but fail to
produce enough flux at the 12 $\mu$m bandpass in the lower
two panels.  Dashed red curves are DL07 dust models with a
constant $U$, selected to be the nearest model match
appropriate to the central star's luminosity and the
nebula's standoff distance, as listed in 
Table~\ref{tab:Tcompare}.  The model curves are normalized
to the 70 $\mu$m point.  The dash-dot green curves  are DL07
dust models having a range of radiation energy density
between $U_{min}$ and $U_{max}$, again normalized to 70
$\mu$m.  Values of $U_{min}$ and $U_{max}$ are labeled
on the plots.  All models have the minimum PAH fraction of
$q_{PAH}$=0.47\% by mass.  Figure~\ref{fig:SED1} shows that
the fixed-$U$ models are generally a poor representation of
the data compared to the variable-$U$ models. The latter,
which include   a contribution from large molecules (PAHs)
provide a better fit to the 12 $\mu$m data than either the
blackbody model or the single-$U$ DL07 models.  Given that
the DL07 models have at least three free parameters
($U_{min}$, $U_{max}$, and PAH fraction) we do not perform a
fit to the data points, but rather we show the model curves
as illustrative examples.  In these four examples, all of
which are hot stars  with large implied radiant energy
density at the location of the nebula, the SEDs are most
consistent with minimal or no PAH content, as might be
expected if shocks act efficiently to destroy  large
molecules.         

\subsection{SEDs of Sources at Shorter Wavelengths}

\begin{figure}
\plotone{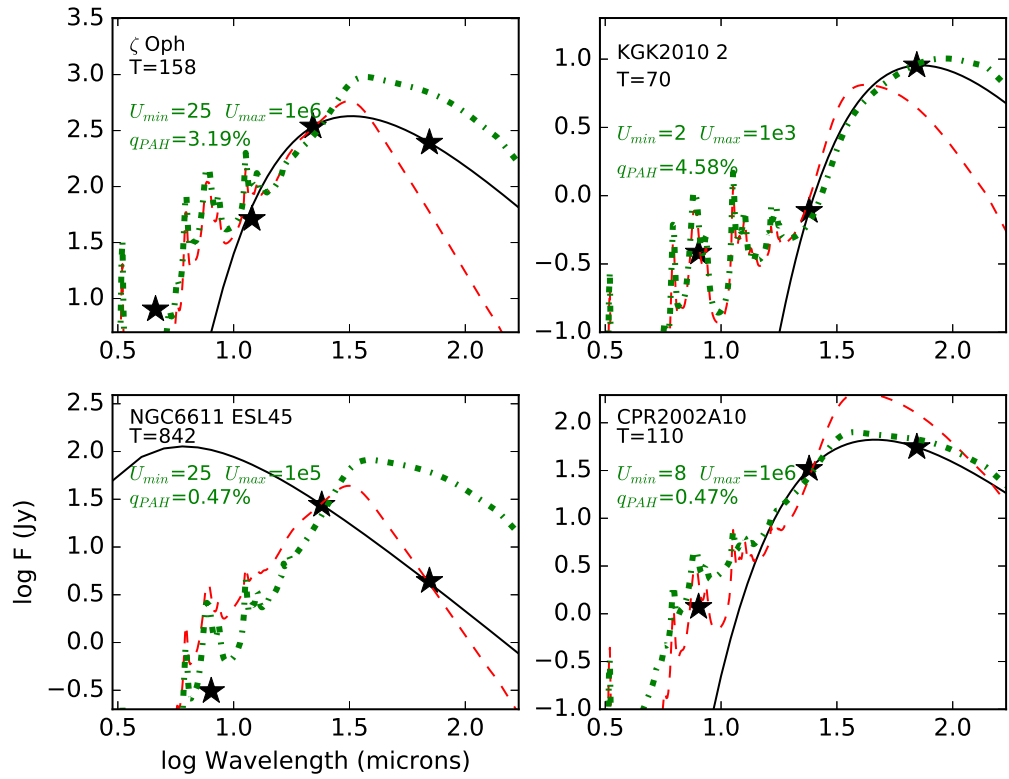}
\caption{Spectral energy distributions for four nebulae from Table~\ref{tab:Tcompare}, including 70 $\mu$m,
24/22 $\mu$m, and the next two shorter wavelength data points.  Notation follows 
Figure~\ref{fig:SED1}. The green dotted curve, depicting one of the DL07 dust models with a range of 
radiant intensity, fit the data better than either the single-intensity models ({\it red}) or the
blackbody fit to the two longest wavelength measurements ({\it solid curve}).
\label{fig:SED2}}
\end{figure}

Additional photometric data at shorter wavelengths can
further help constrain the dust properties.
Figure~\ref{fig:SED2} depicts the spectral energy
distributions for four nebulae ($\zeta$ Oph and KGK2010 2,
as in Figure~\ref{fig:SED1}) plus two additional objects
from Table~\ref{tab:Tcompare}.  The data points include the
70 $\mu$m and 24 or 22 $\mu$m measurements, plus the next
two shorter bandpasses, either $SST$ 5.8 and 8.0 $\mu$m or
$WISE$ 4.5 and 12 $\mu$m.  Models are normalized  to the 24
or 22 $\mu$m data points.  Notations follow those in
Figure~\ref{fig:SED1}.  In three of the four cases the
single-$U$  DL07 dust model of the appropriate $U$ from
Table~\ref{tab:Tcompare} ({\it red}) cannot simultaneously
fit the 70 $\mu$m data point and the shortest wavelength
points.   Similarly, the blackbody curve, in three of the
four cases, significantly under predicts the  flux at the
shortest wavelengths, expected to contain PAH or very hot
dust  contributions.  Generally, the variable-$U$ models
({\it green}) do a better job at matching the short
wavelength data while also over-predicting the 70 $\mu$m
flux.  Notably, the required PAH contribution for KGK2010 2
(upper right panel)  is the maximum included in the DL07
model grid at 4.5\%, as evidenced by the large 8.0 to 24
$\mu$m flux ratio. $\zeta$ Oph is also better with with a
PAH fraction of $q_{PAH}$=3.19\%.  The other two objects seemingly
require no or minimal PAH contribution. NGC6611 ESL 45 is
noteworthy for having the largest 24/70 $\mu$m ratio in our
sample, and correspondingly,  the largest implied color
temperature at 842~K.   The best fitting multi-$U$ models
typically have $U_{min}$=25 and $U_{max}$ of 10$^{3}$  to
10$^{5}$.  For this object, the single-$U$ model provides
the best fit.  

\begin{figure}
\plotone{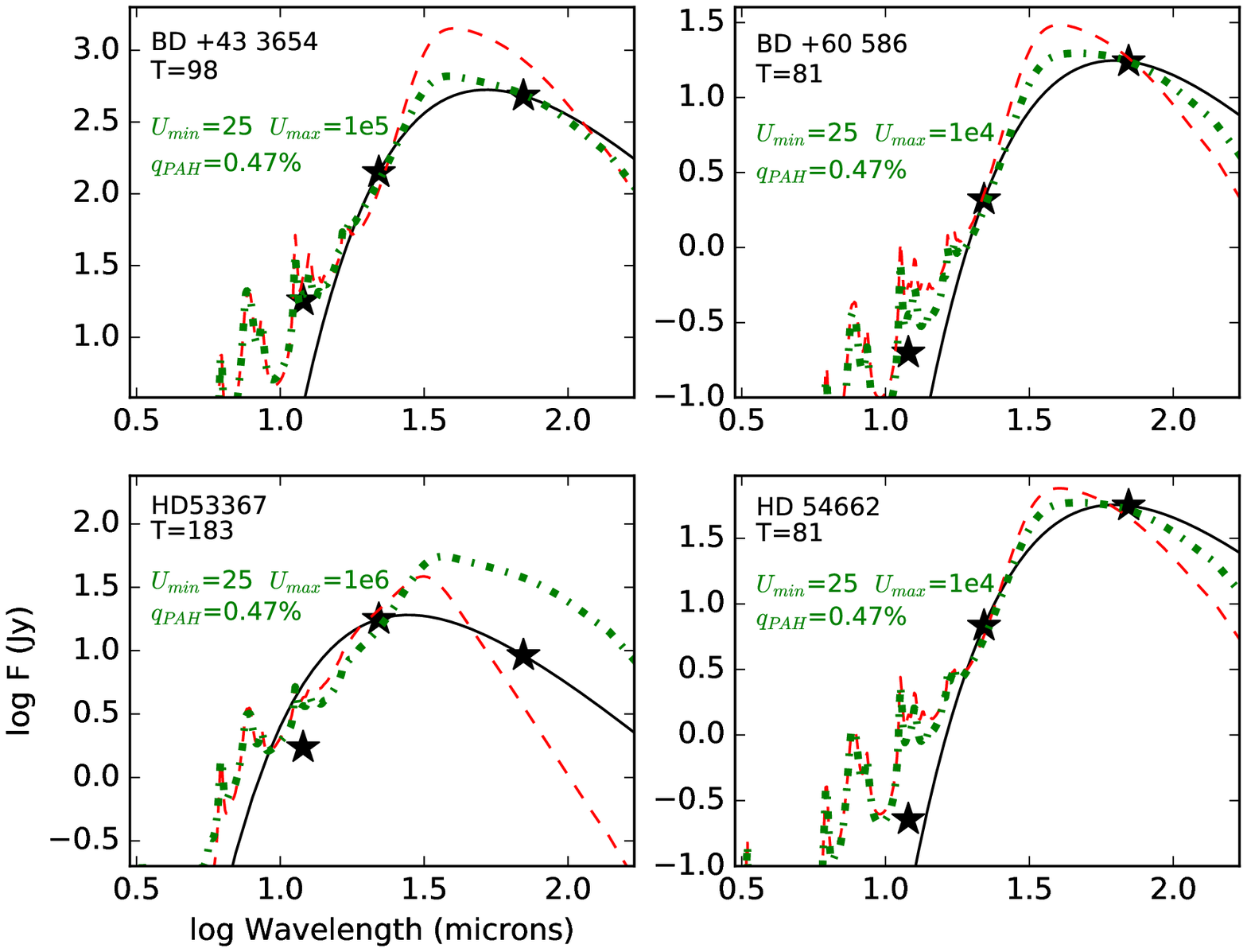}
\caption{Spectral energy distributions for four additional nebulae from Table~\ref{tab:Tcompare},
as in Figure~\ref{fig:SED2}. \label{fig:SED3}}
\end{figure}

Figure~\ref{fig:SED3} show an additional four SEDs from
Table~\ref{tab:Tcompare} with notation as in
Figure~\ref{fig:SED2}.  The best fitting models are similar,
having minimal PAH fraction and $U_{min}$=25 and $U_{max}$ of
10$^{3}$ to 10$^{5}$. The models having a range of radiant 
intensity do a good job of reproducing the data in three of
the four panels.  In the two panels in the right column the
single-$U$ models also reproduce the data as well as the
variable-$U$ models.   In the lower left panel neither of
the DL07 models provide a good fit to the data.  The simple
blackbody fit to the two longest wavelength data points does
a reasonable job of approximating the data for all four
objects.  

\begin{figure}
\plotone{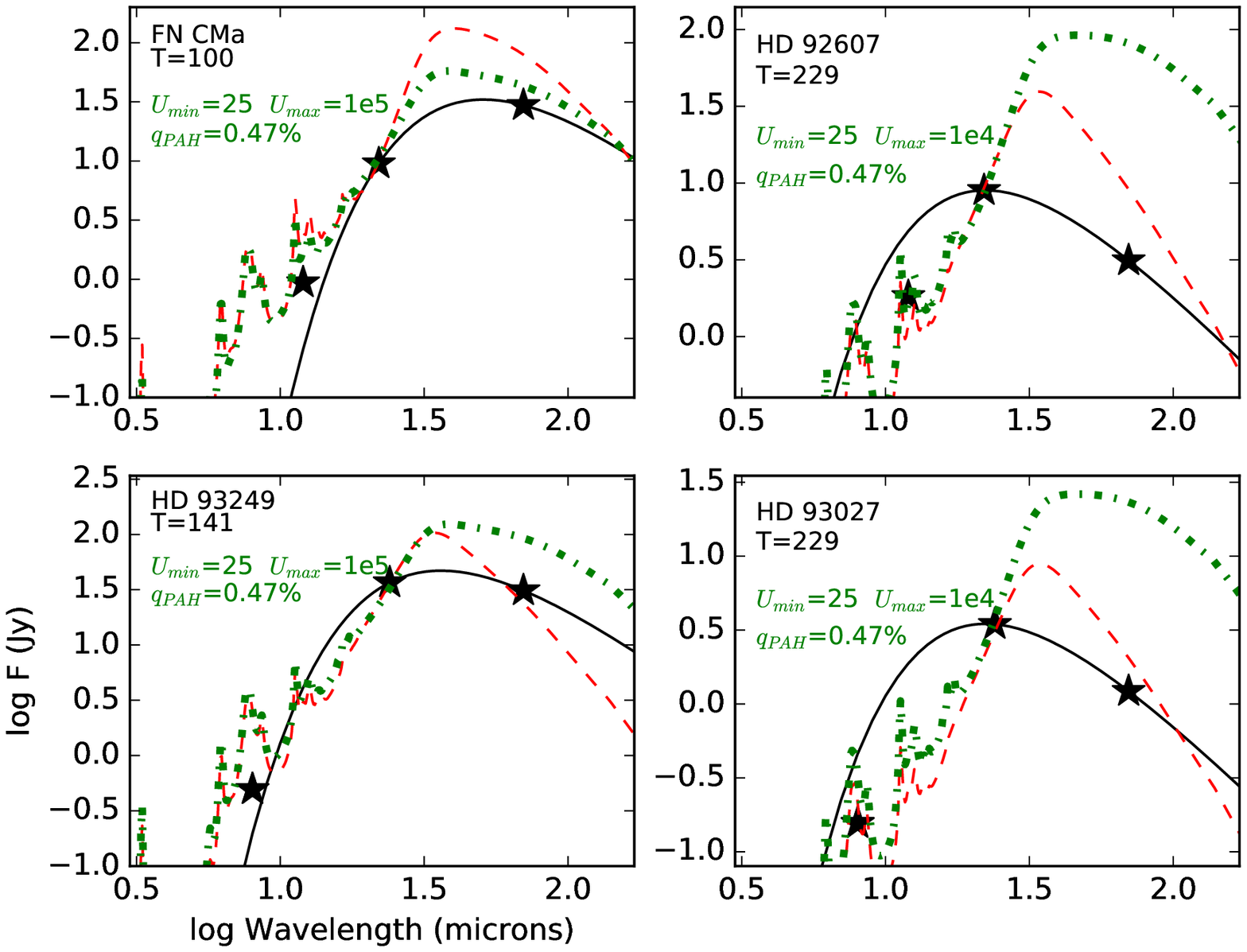}
\caption{Spectral energy distributions for four additional nebulae from Table~\ref{tab:Tcompare},
as in Figure~\ref{fig:SED2}. \label{fig:SED4}}
\end{figure}

Figure~\ref{fig:SED4} show another four SEDs from
Table~\ref{tab:Tcompare}  with notation as in
Figure~\ref{fig:SED2}.  The best fitting variable-$U$ models
are similar, having minimal PAH fraction and $U_{min}$=25 and
$U_{max}$ of 10$^{3}$ to 10$^{5}$.  In three of four cases
the single-$U$ models fit the data better than the
variable-$U$ models.  In general the simple blackbody curves
fit to the two longest wavelength data points are also
reasonable approximations of the overall SEDs.  

\subsection{SEDs of Sources with Strong 8 $\mu$m Detections}

 Although most nebulae are not detected at 8 $\mu$m or
shorter, those that have short-wavelength measurements can
be used to yield insights  regarding the PAH content of the
nebula.   Figure~\ref{fig:sed-PAH1} plots the $SST$ 3.6,
4.5, 5.8, 8.0, and 24 $\mu$m measurements and $HSO$ 70
$\mu$m measurement for four sources having suitable data as
black stars.  The generic source name and fitted 24/70
$\mu$m blackbody ({\it solid black line}) temperature are
labeled in the upper left.    As in previous SED figures,
the green dotted curve  shows a DL07 dust model with a range
of radiant energy density and  the minimum PAH fraction
(0.47\%).  The SED in the upper right panel is G026.1437-0.0420, pictured in Figure~\ref{fig:example1}.
In two of the four panels the green curve  fits
the long wavelength points reasonably well, while in the
other two the blue curve provides a better match to the
data.  In all but the lower left panel the green curve
underpredicts the  flux at the shortest two wavebands.  The
red curve is a model having the same range of $U$ but for a
larger PAH fraction, as labeled in each panel.  In all cases
a very small increase in PAH content is sufficient to 
reproduce the 3.6 and 4.5 $\mu$m fluxes but begins to
overpredict the 8 $\mu$m fluxes.  The blue curve is a model
with a  single radiant energy density ($U$=1$\times$10$^4$,
as typical of stars described in previous subsections) and a
variable PAH content.   This model fits the long wavelength
data better than the variable-$U$ models in two of the four
panels.  It is similar to the red curve in its ability to
predict the short-wavelength data, often overpredicting the
8 $\mu$m flux once the PAH fraction exceeds 1.12\% (the
second lowest model available in the DL07 grid).
Figure~\ref{fig:sed-PAH2} shows SEDs for four additional
sources, as in Figure~\ref{fig:sed-PAH1}. Here, the blue
single-$U$ curve does a better job of reproducing the
long-wavelength data than the variable-$U$ models in all
four panels.   Similar to the four sources in
Figure~\ref{fig:sed-PAH1}, a small PAH fraction of $q_{PAH}$
=0--2\% does a better job of reproducing the
short-wavelength fluxes than the minimum PAH models.  PAH
fractions larger than this overpredict the 8 $\mu$m fluxes,
and in some cases, even the minimum PAH fraction models
overpredict the 8$\mu$m  fluxes.  In seven of the eight
sources the fluxes are monotonically falling at short
wavelengths; only G332.7156+00.5673 shows an upturn, perhaps
due to very hot dust or contamination from
foreground/background stellar sources  along the sightline
to the nebula.  We conclude that, among nebulae with
detections at short wavelengths, single-$U$ models fit the
data better than models with a range of radiation energy
density and models with minimal PAH content are most consistent with the data. 

\section{Inferences Regarding the Ambient ISM Density}

  \citet{Meyer2014} conduct a series of hydrodynamical
simulations of bowshocks generated by main sequence stars
ranging from 10--40 M$_\odot$  and red supergiants from
10--20 M$_\odot$.  Their Figure 24 presents the predicted
correlation between bowshock luminosity and bowshock volume,
in three different tracers, including the infrared continuum
most relevant here.  Figure~\ref{fig:LIR} plots infrared
luminosity  versus standoff distance\footnote{$R_0$ has
been  corrected by a statistical factor of  $\sqrt{2}$ to
account for the effects of the unknown inclination angle on
the measured standoff distance.} cubed, following
\citet{Meyer2014} for the 20 objects in
Table~\ref{tab:Tcompare}. The dashed line is that from 
\citet{Meyer2014} Figure 24 for main sequence star models
and infrared luminosity from reprocessed starlight.  There
is a weak correlation among the data, broadly consistent in
slope with the  models. The data have similar luminosities
to those predicted by \citet{Meyer2014} for 20--40 M$_\odot$
stars,  ranging between 10$^{34}$ and  10$^{37}$ erg
s$^{-1}$.  However, the volumes are about three orders of
magnitude smaller, indicative of much smaller standoff
distances, $R_0$, by about a factor of 10.  $R_0$ scales as 
\citep[e.g.,][]{Wilkin1996},

\begin{equation}
R_0 \propto \sqrt{{{\dot{M} v_w}\over{n_a v_*^2}}  },
\end{equation}

\noindent where $\dot{M}$ is the mass loss rate, $v_W$ is
the stellar wind velocity, $n_a$ is the ambient ISM density,
and $v_*$ is the star's space velocity.   Generally is is
accepted that $v_w$ and $v_*$ are known to factors of two or
better. Some combination of smaller $\dot{M}$ and larger
$n_a$ are apparently needed to reduce $R_0$ by a factor of
10 required to reconcile the data with the models.  

We infer from this discrepancy that  the typical ambient
interstellar densities in our sample may be as much as
factors of 100 larger than the 0.57 $cm^{-3}$ used in the 
\citet{Meyer2014} simulations.  ISM densities larger by
factors of of 3--5 would  be consistent with values of 2--3 
$cm^{-3}$ determined in a few well-studied bowshock nebulae
such as $\alpha$ Orionis \citep{Ueta2008} or $\zeta$ Oph
\citep{Gull1979}.  These objects are nearby, however, and
lie in  relatively unconfused regions of the Plane where
their bowshocks are more readily visible amidst the
unrelated Galactic foreground/background.  The majority of
our sample lies close to the Plane where higher surface
brightnesses are be required for detection, consistent with
the requirement of higher ambient ISM densities.  An
alternative hypothesis requires stellar mass loss rates be
factors of $\sim$100 lower than those adopted by
\citet{Meyer2014}, 10$^{-6.2}$ and 10$^{-7.3}$ M$_\odot$
yr$^{-1}$ for  40 and 20 M$_\odot$ stars, respectively
\citep{Vink2001}.  Some combination of higher ISM density by
factors of up to 100 and lower mass loss rates by factors of
up to 100 would reconcile the data and models in
Figure~\ref{fig:LIR}. 

\section{Conclusions}

This work constitutes the first analysis of the infrared
spectral properties for a large sample of arcuate infrared
nebulae, presumed to be bowshocks in the majority of cases.
Photometry from 3.6 $\mu$m through 160  $\mu$m  provides a
panchromatic characterization of many candidate interstellar
bowshock nebulae powered by early type stars.  

\begin{figure}
\plotone{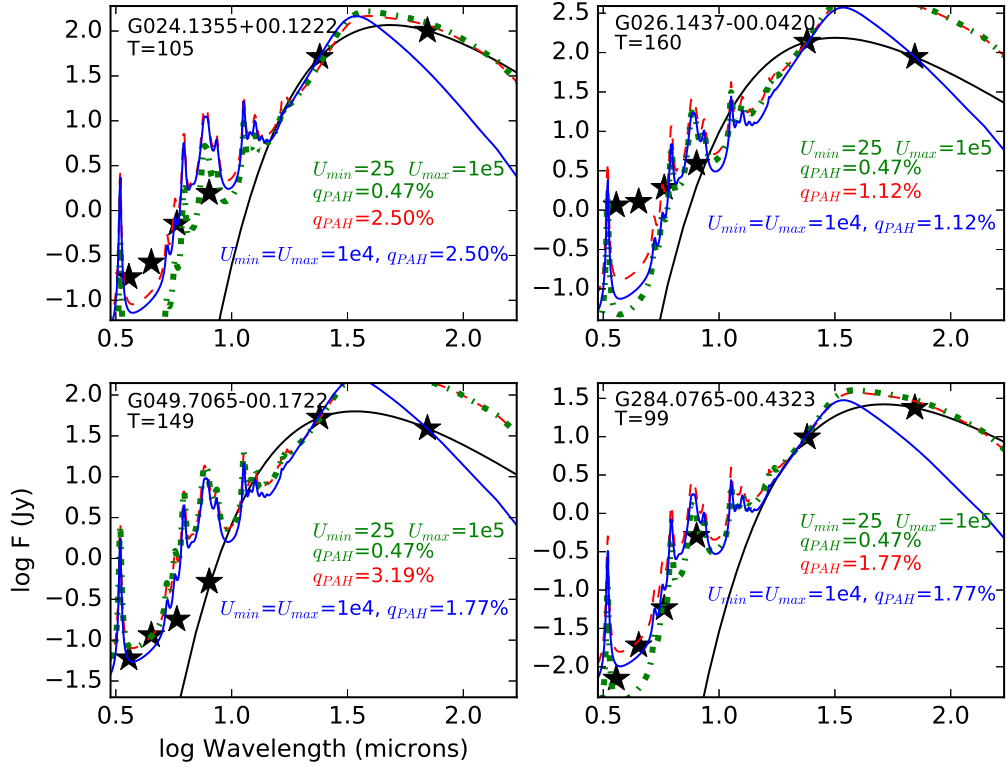}
\caption{Spectral energy distributions for four sources,
labeled at upper left in each panel, having
five-band $SST$ IRAC+MIPS measurements at 3.6--24 $\mu$m and a 
$HSO$ 70 $\mu$m measurement. The black curve is best-fit  blackbody with temperature labeled at upper left.  Colored curves are DL07 interstellar dust models, as described in
the text and notations in each panel. \label{fig:sed-PAH1}}
\end{figure}

\begin{figure}
\plotone{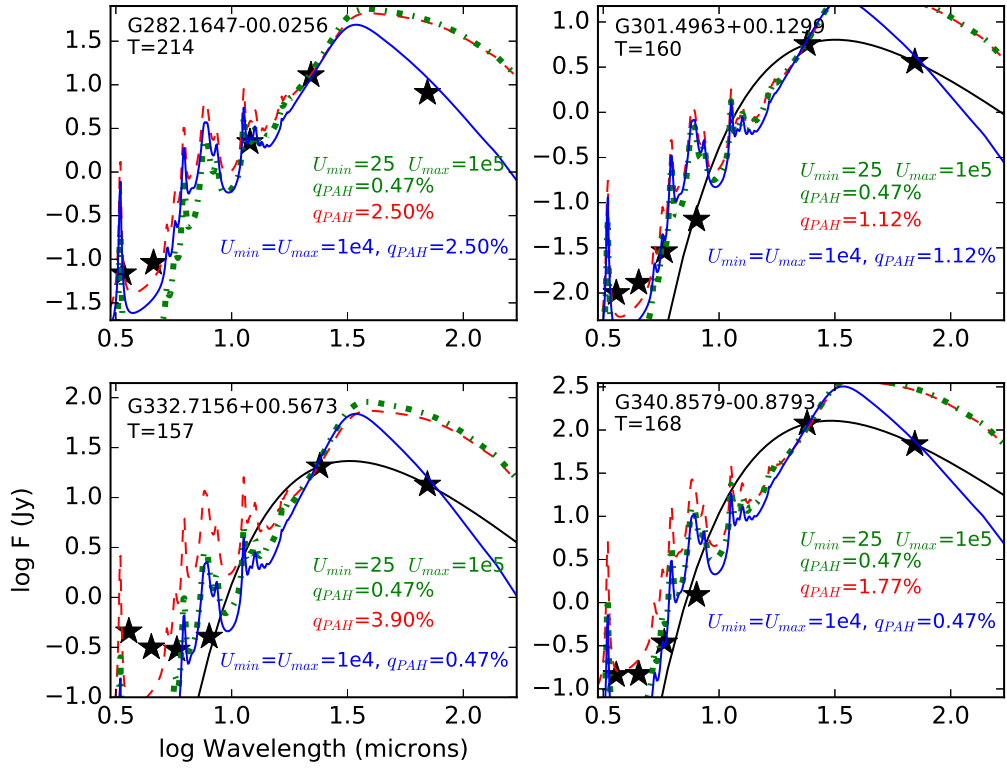}
\caption{Spectral energy distributions for four additional nebulae having short-wavelength
photometry, as in Figure~\ref{fig:sed-PAH1}. \label{fig:sed-PAH2}}
\end{figure}

\begin{figure}
\plotone{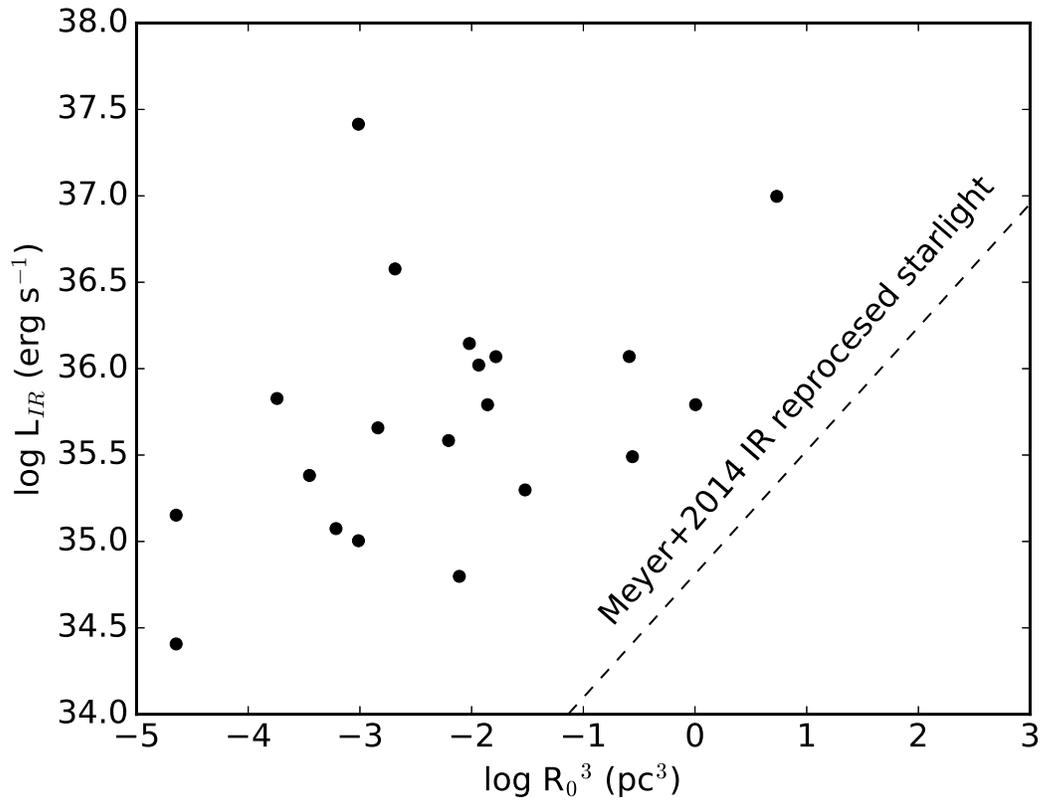}
\caption{Infrared luminosity versus the cube of the bowshock
standoff distance, following Figure 24 of \citet{Meyer2014}. Points are the measurements
of 20 bowshocks having suitable phtometry and distance data.  The dashed line
shows the predicted correlation from their hydrodynamical simulations of main sequence
stars. \label{fig:LIR}}
\end{figure}

\begin{enumerate}

\item Essentially all nebulae identified at the $WISE$ 22
$\mu$m bandpass are also detected at the $WISE$ 12 $\mu$m
band.   The majority of objects identified at 24/22 $\mu$m 
$SST$/$WISE$ bandpasses have 70 $\mu$m $HSO$ detections (422
objects,; 60\%), but only a small fraction (39 objects;
$<$6\%) are detected at 160 $\mu$m.  This low detection rate
is a consequence of being on the Rayleigh-Jeans tail of the
blackbody spectrum and the complex emission from
foreground/background structures in the Galactic Plane.

\item Only a small fraction (178 objects; 25\%) are detected
at $SST$ 8 $\mu$m, where PAH emission often dominates the
emission, or at shorter wavelengths.  This suggests that PAH
contributions are not generally present in these nebulae, a
conclusion supported by comparing ISM dust models to a
subset of well-measured objects.  

\item  Spectral energy distributions  peak in the 12--70
$\mu$m range, in agreement with the hydrodynamical
simulations of \cite{Meyer2016}. 

\item Color temperatures derived from 70/160 $\mu$m ratios
are systematically cooler than those from 24/70  $\mu$m
ratios. We infer from this that either the 160 $\mu$m
measurements are contaminated by unrelated
background/foreground emission (appears likely in some
cases), that the nebulae are intrinsically multi-temperature
dust structures, as might be expected if a bowshock is being
driven into a higher density structure such as a molecualr
cloud, or that the 24/22 $\mu$m bandpasses contain
disproportionate contribution from very small,
stochastically heated dust grains.  

\item Among a subset of 20 objects with well-determined
distances, sizes, stellar effective temperatures, and
stellar radii, we find that the radiant energy density from
the star at the location of the nebula is 10$^2$ to 10$^5$
larger than the ambient interstellar radiation field,
indicating that the early type stars dominate the energetics
of the nebulae.   Both of the infrared color temperatures
are hotter, by 76\% on average,  than the nominal steady
state dust temperature if the dust were in radiative
equilibrium, consistent with shocks or additional heating
sources in these nebulae.  

\item  Among the 20 well-characterized objects, we find  no
correlation between the  ratio of stellar-to-nebular
luminosity and the standoff distance, $R_0$. There is a weak
correlation between infrared luminosity and $R_0^3$, as
predicted by numerical simulations of \citet{Meyer2014}.
However, the data are offset toward smaller $R_0$ compared to
the simulations. We infer that ambient ISM densities in the
vicinity of bowshocks are are higher by factors of $\sim$100
or stellar mass loss rates are lower by factors of
$\sim$100, or some combination thereof, relative to the
values adopted in the simulations.

\item Analysis of the SEDs for 20 well-characterized objects
shows that the shortest wavebands (8 or 12 $\mu$m) rarely
have  fluxes large enough to require significant PAH
contributions. Most SEDs are consistent with DL07 dust
spectra having a constant radiant energy density (single-$U$
models) or a  range of radiant intensity from $U=$25 to
$U=10^3$ or $10^4$ times the ambient interstellar radiation
field for a minority of the objects.  In about half the cases the single-$U$ models
provide better fits to the 8--70 $\mu$m data, while the
variable-$U$ models appear to provide better fits to the
small number of objects having 24--160 $\mu$m photometry.  
In nearly all cases, the models with minimal PAH content
provide the best fits, indicating that PAH contribution to
most candidate bowshock nebulae is small.  Even in sources
with strong 8 $\mu$m detections (Figures~\ref{fig:sed-PAH1}
and \ref{fig:sed-PAH2})   a small increase in the PAH
fraction to $q_{PAH}$=1--2\% appears sufficient to reproduce
the short-wavelength 3.6--5.8 $\mu$m data.  This is
substantially smaller than the median $q_{PAH}$=4.1\% for
nearby galaxies \citep{Dale2017} and does not vary, on
average, by more than half a percent as a function of
galactocentric radius, with $q_{PAH}$ dropping slightly at smaller radii \cite{Sandstrom2013}.  Taken together,
these trends are consistent with shocks or UV radiation
fragmenting large dust grains and large molecules within the
nebulae to the point where PAHs are not a dominant component
of the infrared SEDs.

\end{enumerate}

Additional spectroscopic observations, both optical and
infrared, are needed to characterize the central stars of
these nebulae and enable a more comprehensive analysis  of
the energy balance and physical properties in a large sample
of arcuate candidate-bowshock nebulae.   Because the
radiation field is convincingly dominated by a single hot
star, these objects make enticing laboratories for studying
the effects of UV radiation on dust properties and for
testing models of dust emission against the data.  
Sensitive optical or infrared spectroscopy may be able to
detect diagnostic  fine-structure or forbidden lines which
could be used as shock tracers or density  indicators. 
Unfortunately the surface brightnesses of the nebulae are
all below the detection thresholds of airborne infrared
instruments on SOFIA, but some objects  are good targets for
the new suite of infrared instruments on the upcoming {\it
James Webb Space Telescope}.  Mid-infrared spectra  would
help constrain the grain size distribution, providing
evidence for  shock-induced  destruction of large grains and
PAHs.  Spatially resolved infrared spectroscopy has the
potential to measure variations in  dust temperature,
composition, and ionization structure across the putative
shock fronts. Our team's ongoing work  will provide more
precise characterizations of the central stars driving
bowshock nebulae, including improved spectral
classifications for a larger sample, measurements of their
space velocities relative to the local interstellar medium,
and the possible multiplicity among this sample of early
type stars. 

\acknowledgments This work has been supported by the
National Science Foundation through grants AST-1412845 and 
AST-1560461 (REU). We thank Karin Sandstrom, Danny Dale, and
the anonymous reviewer for  suggestions that substantially
contributed to this work.

\vspace{5mm}
\facilities{SST, WISE, HSO, HIPPARCOS}

\newpage

\begin{deluxetable}{rrrrrrrrrrrrrrrr}
\tablecaption{$SST$ and $HSO$ Photometry of Candidate Bowshock Nebulae \label{tab:Phot1}}
\tabletypesize{\scriptsize} 
\rotate
\tablehead{\colhead{ID} &\colhead{Name} &\colhead{R.A.}  &\colhead{Decl.} &\colhead{Radius} &\colhead{Height}     &\colhead{F$_{3.6}$} &\colhead{F$_{4.5}$} &\colhead{F$_{5.8}$}&\colhead{F$_{8.0}$}&\colhead{F$_{24}$} &\colhead{F$_{70}$} &\colhead{F$_{160}$} &\colhead{Peak$_{70}$}        &\colhead{T$_{24/70}$}&\colhead{T$_{70/160}$} \\
            \colhead{}  &\colhead{}     &\colhead{(2000)}&\colhead{(2000)}&\colhead{(arcsec)} &\colhead{(arcsec)} &\colhead{(Jy)}      &\colhead{(Jy)}      &\colhead{(Jy)}     &\colhead{(Jy)}     &\colhead{(Jy)}	&\colhead{(Jy)}     &\colhead{(Jy)}	 &\colhead{(Jy arcmin$^{-2}$)}   &\colhead{(K)}  &\colhead{(K)} \\
            \colhead{(1)}  &\colhead{(2)}     &\colhead{(3)}&\colhead{(4)}&\colhead{(5)} &\colhead{(6)}           &\colhead{(7)}      &\colhead{(8)}      &\colhead{(9)}     &\colhead{(10)}     &\colhead{(11)}	&\colhead{(12)}     &\colhead{(13)}	 &\colhead{(14)}   &\colhead{(15)}  &\colhead{(16)} 
	    }
\colnumbers
\startdata
1 & G000.1169-00.5703 & 17 48 07.70 & -29 07 55.5 & 38 & 44 & -0.074 & -0.150 & -0.086 & 0.255 & 8.730 & 28.300 & -22.000 & 126.5 & 92 & -99 \\
2 & G000.3100-01.0495 & 17 50 27.59 & -29 12 46.8 & 16 & 25 & -0.258 & -0.149 & -0.114 & 0.041 & 0.453 & 0.450 & -2.000 & 24.6 & 132 & -99 \\
3 & G001.0563-00.1499 & 17 48 41.78 & -28 06 37.8 & 13 & 21 & -0.139 & -0.120 & -0.230 & 0.710 & 13.305 & 39.000 & -20.000 & 372.6 & 95 & -99 \\
4 & G001.2588-00.0780 & 17 48 53.40 & -27 53 59.7 & 17 & 19 & -0.065 & -0.375 & -0.062 & 0.108 & 1.023 & 3.910 & -7.000 & 355.0 & 89 & -99 \\
5 & G003.5118-00.0470 & 17 53 56.10 & -25 56 50.3 & 12 & 23 & -0.055 & -0.038 & -0.020 & -0.010 & 0.273 & -0.400 & -3.800 & -99.9 & -99 & -99 \\
6 & G003.7391+00.1425 & 17 53 43.26 & -25 39 19.2 & 11 & 17 & -0.086 & -0.061 & -0.059 & -0.025 & 0.370 & 1.300 & -6.400 & 56.2 & 90 & -99 \\
7 & G003.8417-01.0440 & 17 58 30.64 & -26 09 49.1 & 13 & 13 & 0.103 & 0.106 & 0.496 & 1.311 & 0.830 & 21.700 & 10.600 & 179.3 & 61 & 80 \\
8 & G004.3087+00.2222 & 17 54 41.61 & -25 07 25.6 & 10 & 12 & -0.083 & -0.250 & -0.056 & -0.093 & 0.609 & 4.700 & 2.400 & 59.8 & 76 & 77 \\
9 & G004.7315-00.3875 & 17 57 57.62 & -25 03 53.6 & 9 & 10 & -0.017 & -0.015 & -0.017 & -0.015 & 0.391 & 3.300 & -4.800 & 45.7 & 75 & -99 \\
10 & G004.8449-00.9309 & 18 00 17.51 & -25 14 14.7 & 9 & 13 & -0.195 & -0.114 & -0.079 & 0.054 & 0.259 & -0.100 & -2.300 & -99.9 & -99 & -99 \\
11 & G005.5941+00.7335 & 17 55 36.25 & -23 45 21.8 & 9 & 11 & -0.116 & -0.076 & -0.032 & 0.086 & 0.758 & 0.500 & -0.200 & 14.1 & 158 & -99 \\
12 & G005.6985-00.6350 & 18 01 06.15 & -24 21 34.4 & 16 & 19 & -0.048 & -0.030 & -0.052 & -0.116 & 1.795 & -1.100 & -3.300 & -99.9 & -99 & -99 \\
14 & G006.2977-00.2012 & 18 00 40.59 & -23 36 50.7 & 17 & 15 & -0.089 & -0.064 & -0.029 & 0.060 & 1.065 & 7.500 & -1.100 & 123.0 & 77 & -99 \\
15 & G006.3600-00.1846 & 18 00 44.92 & -23 33 06.3 & 7 & 10 & -0.034 & -0.188 & -0.024 & -0.054 & 0.318 & -5.700 & -1.000 & -99.9 & -99 & -99 \\
16 & G006.8933+00.0743 & 18 00 55.27 & -22 57 36.9 & 14 & 12 & -0.080 & -0.053 & -0.101 & -0.122 & 2.274 & 7.100 & -1.000 & 80.8 & 93 & -99 \\
17 & G007.5265-00.2652 & 18 03 33.49 & -22 34 38.3 & 4 & 6 & -0.109 & -0.072 & -0.024 & -0.019 & 0.266 & -0.400 & -1.300 & -99.9 & -99 & -99 \\
18 & G008.3690+00.0239 & 18 04 15.53 & -21 42 05.3 & 27 & 16 & -0.164 & -0.131 & -0.131 & 0.189 & 3.254 & 1.800 & -6.400 & 112.5 & 172 & -99 \\
19 & G009.0177+00.1410 & 18 05 11.21 & -21 04 43.3 & 15 & 17 & -0.052 & -0.033 & -0.030 & -0.005 & 0.226 & -0.330 & -2.000 & -99.9 & -99 & -99 \\
20 & G009.6852-00.2025 & 18 07 51.86 & -20 39 48.8 & 7 & 10 & -0.026 & -0.017 & -0.010 & -0.027 & 0.111 & -0.070 & -0.700 & -99.9 & -99 & -99 \\
21 & G010.1395-00.0350 & 18 08 10.89 & -20 11 06.4 & 21 & 17 & -0.038 & -0.033 & -0.157 & -0.496 & 1.137 & 3.700 & -1.900 & 84.4 & 92 & -99 \\
\enddata
\tablecomments{Typical flux uncertainties are 25\% in each band. 
Negative flux values indicate non-detections with 1$\sigma$ upper limits given by -1$\times$ the listed value. 
(1) ID from \citet{Kobulnicky2016}; (2) generic identifier by Galactic coordinates; (3) Right Ascension; (4) Declination;
(5) nebula angular radius in arcsec; (6) nebula angular height in arcsec; (7--11) flux in Jy at $SST$ 3.6/4.5/5.8/8.0/24 $\mu$m bands; (12--13) flux in Jy at $HSO$ 70/160 $\mu$m bands; 
(14) peak surface brightness above local background at 70 $\mu$m in Jy arcmin$^{-2}$; (15--16) color temperature derived from the 24/70 $\mu$m and 70/160 $\mu$m
measurements, respectively. }
\end{deluxetable}
\newpage

\begin{deluxetable}{rrrrrrrrrrrrrrr}
\tablecaption{$WISE$ and $HSO$ Photometry of Candidate Bowshock Nebulae \label{tab:Phot2}}
\tabletypesize{\scriptsize} 
\rotate
\tablehead{\colhead{ID} &\colhead{Name} &\colhead{R.A.}  &\colhead{Decl.} &\colhead{Radius} &\colhead{Height}     &\colhead{F$_{3.4}$} &\colhead{F$_{4.6}$} &\colhead{F$_{12}$} &\colhead{F$_{22}$} &\colhead{F$_{70}$} &\colhead{F$_{160}$} &\colhead{Peak$_{70}$} &\colhead{T$_{22/70}$}&\colhead{T$_{70/160}$} \\
           \colhead{}   &\colhead{    } &\colhead{(2000)}&\colhead{(2000)}&\colhead{(arcsec)}&\colhead{(arcsec)}  &\colhead{(Jy)}      &\colhead{(Jy)}     &\colhead{(Jy)}     &\colhead{(Jy)}     &\colhead{(Jy)}     &\colhead{(Jy)}      &\colhead{(Jy arcsec$^{-2}$) }  &\colhead{(K)}  &\colhead{(K)}  \\
            \colhead{(1)}  &\colhead{(2)}     &\colhead{(3)}&\colhead{(4)}&\colhead{(5)}      &\colhead{(6)}           &\colhead{(7)}      &\colhead{(8)}      &\colhead{(9)}     &\colhead{(10)}     &\colhead{(11)}	&\colhead{(12)}     &\colhead{(13)}	 &\colhead{(14)}   &\colhead{(15)}  
}
\colnumbers
\startdata
13 & G006.2812+23.5877 & 16 37 09.54 & -10 34 01.5 & 173 & 404 & 6.250 & 7.995 & 51.320 & 343.77 & 246.700 & 61.550 & 9.8 & 158 & 240 \\
39 & G013.4900+01.6618 & 18 08 46.51 & -16 25 56.9 & 108 & 92 & -0.675 & -0.677 & 3.406 & 21.43 & -99.999 & -99.999 & -99.9 & -99 & -99 \\
54 & G015.0813-00.6570 & 18 05 58.84 & -14 11 53.0 & 101 & 91 & -0.113 & -0.472 & 3.298 & 28.89 & -99.999 & -99.999 & -99.9 & -99 & -99 \\
64 & G016.8993-01.1152 & 18 25 38.90 & -14 45 05.7 & 53 & 39 & -0.265 & -0.403 & 4.194 & 20.76 & -99.999 & -99.999 & -99.9 & -99 & -99 \\
66 & G016.9848+01.7482 & 18 15 23.97 & -13 19 35.3 & 102 & 127 & 0.368 & 0.509 & 9.384 & 52.45 & -99.999 & -99.999 & -99.9 & -99 & -99 \\
154 & G031.1096+03.6457 & 18 35 08.21 & +00 02 34.8 & 30 & 28 & 0.053 & 0.048 & 0.184 & 0.79 & -99.999 & -99.999 & -99.9 & -99 & -99 \\
240 & G045.9397+03.2506 & 19 03 40.72 & +13 03 11.5 & 17 & 18 & -0.010 & -0.005 & -0.005 & 0.01 & -99.999 & -99.999 & -99.9 & -99 & -99 \\
272 & G050.9339+03.0747 & 19 13 48.34 & +17 24 15.5 & 16 & 21 & -0.008 & -0.004 & -0.005 & 0.03 & -99.999 & -99.999 & -99.9 & -99 & -99 \\
298 & G059.9225-01.9671 & 19 51 08.28 & +22 49 53.9 & 13 & 13 & 0.024 & 0.020 & 0.037 & 0.14 & -99.999 & -99.999 & -99.9 & -99 & -99 \\
310 & G061.8355+02.9452 & 19 36 31.44 & +26 56 30.2 & 25 & 25 & -0.015 & -0.008 & -0.001 & 0.06 & -99.999 & -99.999 & -99.9 & -99 & -99 \\
313 & G063.5153-01.4433 & 19 57 17.76 & +26 10 49.2 & 13 & 23 & -0.007 & -0.004 & 0.035 & 0.02 & -99.999 & -99.999 & -99.9 & -99 & -99 \\
315 & G064.0602+01.6348 & 19 46 39.32 & +28 13 23.2 & 35 & 73 & -0.077 & -0.009 & -0.007 & 0.25 & -99.999 & -99.999 & -99.9 & -99 & -99 \\
316 & G064.7582+00.2889 & 19 53 30.15 & +28 08 20.1 & 8 & 10 & -0.001 & -0.003 & -0.008 & 0.06 & -99.999 & -99.999 & -99.9 & -99 & -99 \\
317 & G067.1370-00.6744 & 20 02 57.58 & +29 39 49.9 & 29 & 24 & -0.008 & -0.005 & 0.031 & 0.03 & -99.999 & -99.999 & -99.9 & -99 & -99 \\
318 & G073.2946-01.6939 & 20 22 58.66 & +34 14 20.3 & 21 & 19 & -0.001 & -0.002 & 0.016 & 0.01 & -99.999 & -99.999 & -99.9 & -99 & -99 \\
319 & G073.3536+02.5872 & 20 05 38.06 & +36 39 37.3 & 5 & 151 & -0.002 & -0.002 & -0.020 & 0.13 & -99.999 & -99.999 & -99.9 & -99 & -99 \\
320 & G073.6200+01.8522 & 20 09 25.91 & +36 29 19.1 & 37 & 24 & -0.056 & -0.065 & 0.089 & 2.98 & 6.480 & -2.610 & 7.0 & 110 & -99 \\
321 & G074.1929+00.9964 & 20 14 33.30 & +36 29 48.6 & 8 & 9 & -0.001 & -0.020 & 0.021 & 0.12 & -99.999 & -99.999 & -99.9 & -99 & -99 \\
322 & G074.3117+01.0041 & 20 14 51.09 & +36 35 59.7 & 21 & 25 & -0.002 & -0.001 & 0.020 & 0.51 & 1.600 & -1.020 & 3.5 & 100 & -99 \\
323 & G075.1730-00.5964 & 20 23 50.71 & +36 24 26.4 & 13 & 7 & 0.023 & 0.031 & 0.155 & 0.89 & -99.999 & -99.999 & -99.9 & -99 & -99 \\
\enddata
\tablecomments{Typical flux uncertainties are 25\% in each band. Negative flux values indicate non-detections with 1$\sigma$ upper limits given by -1$\times$ 
the listed value. (1) ID from \citet{Kobulnicky2016}; (2) generic identifier by Galactic coordinates; (3) Right Ascension; (4) Declination;
(5) nebula angular radius in arcsec; (6) nebula angular height in arcsec; (7--10) flux in Jy at $WISE$ 3.4/4.6/12/22 $\mu$m bands; (11--12) flux in Jy at $HSO$ 70/160 $\mu$m bands; 
(13) peak surface brightness above local background at 70 $\mu$m in Jy arcmin$^{-2}$; (14--15) color temperature derived from the 22/70 $\mu$m and 70/160 $\mu$m
measurements, respectively.}
\end{deluxetable}
\newpage

\begin{deluxetable}{c|rrr}
\tablecaption{$SST$+$HSO$ Detections by Band\label{tab:freqs1}}
\tablehead{
\colhead{Band} & \colhead{$SST$ 24} & \colhead{$HSO$ 70} & \colhead{$HSO$ 160}}
\startdata
24  & 617 & 399 & 35 \\
8   & 106 & 90 & 20 \\
3.6 &  32 & 26 & 8 \\
\enddata
\end{deluxetable}

\begin{deluxetable}{c|rrr}
\tablecaption{$WISE$+$HSO$ Detections by Band\label{tab:freqs2}}
\tablehead{
\colhead{Band} & \colhead{$WISE$ 22} & \colhead{$HSO$ 70} & \colhead{$HSO$ 160}
}
\startdata
22  &  85 &  23 &  4 \\
12  &  72 &  23 &  4 \\
3.4 &  12 &   2 &  1 \\
\enddata
\end{deluxetable}

\newpage

\begin{deluxetable}{rcrcrrrrrrrrrcr}
\tablecaption{Derived Parameters for Stars with Known Characteristics \label{tab:Tcompare}}
\tablehead{
\colhead{ID} & \colhead{Name} & \colhead{Alt. name} & \colhead{Sp.T.} & \colhead{T$_{eff}$} & \colhead{R$_*$ }     & \colhead{D} & \colhead{$R_0$ }    & \colhead{$R_0$}  & \colhead{$U$} & \colhead{$T_{SS}$}& \colhead{$T_{22/70}$} & \colhead{$T_{70/160}$} &\colhead{$F_{IR}$}  & \colhead{$L_{*}/L_{IR}$} \\
\colhead{}   & \colhead{    } & \colhead{         } & \colhead{     } & \colhead{(K)}       & \colhead{($R_\odot$)}& \colhead{(kpc)} & \colhead{(\arcsec)} & \colhead{(pc)}   & \colhead{}    & \colhead{(K)}     & \colhead{(K)}	     & \colhead{(K)}	 & \colhead{(erg s$^{-1}$ cm$^{-2}$)} & \colhead{} \\
\colhead{(1)}   & \colhead{(2)} & \colhead{(3)} & \colhead{(4)} & \colhead{(5)}       & \colhead{(6)}& \colhead{(7)} & \colhead{(8)} & \colhead{(9)}   & \colhead{(10)}    & \colhead{(11)}     & \colhead{(12)}	     & \colhead{(13)}	 & \colhead{(14)} & \colhead{(15)} 
}
\startdata
 13 & $\zeta$ Oph    & G006.2812+23.5877 & O9.2IV   & 31000 & 7.2 & 0.14 &29  & 0.02 & 1.7$\times10^5$ & 123 & 158 & 240      & 6.4$\times10^{-8}$ & 1110 \\
 67 & NGC 6611 ESL 45& G017.0826+00.9744 & O9V      & 31500 & 7.7 & 1.99 &7.5 & 0.07 & 1.6$\times10^4$ &  82 & 842 & \nodata  & 5.8$\times10^{-8}$ & 8.4  \\
329 & KGK 2010 10    & G077.0505$-$00.6094 & O7V    & 35500 & 9.3 & 1.32 &10  & 0.06 & 4.6$\times10^4$ &  98 & 131 & \nodata  & 5.1$\times10^{-10}$ & 4400  \\
331 & LS II +39 53   & G078.2869+00.7780 & O7V      & 35500 & 9.3 & 1.32 &25  & 0.18 & 7.3$\times10^3$ &  72 & 195 & \nodata  & 5.3$\times10^{-9}$ & 430  \\
338 & CPR2002A10     & G078.8223+00.0959 & O9V      & 31500 & 7.7 & 1.32 &23  & 0.15 & 3.8$\times10^3$ &  65 & 110 & \nodata  & 6.9$\times10^{-9}$ & 160  \\
339 & CPR2002A37     & G080.2400+00.1354 & O5V      & 41500 & 11.1& 1.32 &70  & 0.45 & 2.6$\times10^3$ &  61 & 120 & \nodata  & 6.0$\times10^{-9}$ & 1100  \\
341 & KGK2010 1      & G080.8621+00.9749 & B2V      & 20900 & 5.4 & 1.32 &20  & 0.13 & 4.7$\times10^2$ &  46 & 126 & \nodata  & 1.9$\times10^{-9}$ & 43  \\
342 & KGK2010 2      & G080.9020+00.9828 & B2V      & 20900 & 5.4 & 1.32 &10  & 0.06 & 1.9$\times10^3$ &  58 & 70  & 69       & 6.0$\times10^{-10}$ & 140  \\
344 & BD +43 3654    & G082.4100+02.3254 & O4If     & 40700 & 19  & 1.32 &193 & 1.24 & 9.1$\times10^2$ &  51 & 98  & 174      & 5.0$\times10^{-8}$ & 350  \\
368 & KM Cas         & G134.3552+00.8182 & O9.5V    & 30500 & 7.4 & 2.00 &14  & 0.13 & 3.7$\times10^3$ &  64 & 96  & \nodata  & 1.4$\times10^{-10}$ & 2400  \\
369 & BD +60 586     & G137.4203+01.2792 &O7.5V/O8III&34400 & 8.9 & 2.00 &73  & 0.71 & 3.2$\times10^2$ &  43 & 81  & 148      & 1.4$\times10^{-9}$ & 640  \\
380 & HD 53367       & G223.7092$-$01.9008 & B0IVe  & 30000 & 7.4 & 0.26 &15  & 0.02 & 1.8$\times10^5$ & 124 & 183 & \nodata  & 3.3$\times10^{-9}$ & 5400  \\
381 & HD 54662      & G224.1685$-$00.7784 & O7V    & 35500 & 9.4 & 0.63 &71  & 0.22 & 4.2$\times10^3$ &  66 & 82  & \nodata  & 4.4$\times10^{-9}$ & 2300  \\
382 & FN CMa         & G224.7096$-$01.7938 & B2Ia   & 17600 & 30  & 0.93 &101 & 0.45 & 5.8$\times10^2$ &  47 & 100 & \nodata  & 3.2$\times10^{-9}$ & 980  \\
406 & HD 92607       & G287.1148$-$01.0236 & O8V+O9V& 33400 & 8.5 & 2.30 &16  & 0.18 & 4.1$\times10^3$ &  66 & 230 & \nodata  & 1.9$\times10^{-9}$ & 300  \\
407 & HD 93249       & G287.4071$-$00.3593 & O9III  & 33400 & 13.6& 2.30 &7.8 & 0.09 & 3.1$\times10^4$ &  92 & 141 & \nodata  & 6.3$\times10^{-9}$ & 160  \\
409 & HD 93027       & G287.6131$-$01.1302 & O9.5IV & 30500 & 7.4 & 2.30 &7.4 & 0.08 & 9.9$\times10^3$ &  76 & 229 & \nodata  & 7.6$\times10^{-10}$ & 330  \\
410 & HD 305536      & G287.6736$-$01.0093 & O9.5V+?& 30500 & 7.4 & 2.30 &3.7 & 0.04 & 3.9$\times10^4$ &  96 & 102 & \nodata  & 1.1$\times10^{-9}$ & 220  \\
411 & HD 305599      & G288.1505$-$00.5059 & B0Ib   & 25000 & 30  & 2.30 &4.2 & 0.05 & 2.3$\times10^5$ & 128 & 106 & \nodata  & 4.0$\times10^{-10}$ & 5100  \\
413 & HD 93683       & G288.3138$-$01.3085 & O9V    & 31500 & 7.7 & 2.30 &15  & 0.17 & 2.9$\times10^3$ &  62 & 123 & \nodata  & 1.0$\times10^{-9}$ & 360  \\
\enddata
\end{deluxetable}

\newpage

\newpage

\newpage

\newpage

\newpage

\newpage

\newpage

\newpage

\newpage

\newpage

\newpage

\newpage

\clearpage
\newpage

\newpage

\newpage

\newpage

\end{document}